\PassOptionsToPackage{table, xcdraw, dvipsnames}{xcolor}
\documentclass{article}

\usepackage{arxiv}
\usepackage{setspace}
\usepackage[toc,page]{appendix}
\usepackage{amsmath}
\usepackage{ amssymb }
\usepackage{natbib}
\usepackage{subfig}
\usepackage{dsfont}
\usepackage{graphicx}
\usepackage[utf8]{inputenc} % allow utf-8 input
\usepackage[T1]{fontenc}    % use 8-bit T1 fonts
\usepackage{hyperref}       % hyperlinks
\usepackage{url}            % simple URL typesetting
\usepackage{booktabs}       % professional-quality tables
\usepackage{amsfonts}       % blackboard math symbols
\usepackage{nicefrac}       % compact symbols for 1/2, etc.
\usepackage{microtype}      % microtypography
\usepackage{tikz}
\usepackage{cancel}
\usepackage{wasysym}
\usepackage{scalerel}
\usepackage{multirow}
\usepackage{adjustbox}
\newcolumntype{R}[2]{%
    >{\adjustbox{angle=#1,lap=\width-(#2)}\bgroup}%
    l%
    <{\egroup}%
}
\newsavebox{\mytikzbox}
\savebox{\mytikzbox}{%
  \begin{tikzpicture}[scale=0.2] % Adjust scale as needed
    \draw (0, 0) rectangle (1, 1);
    \draw (0.0,0.0) -- (1,1);
    \draw (0,1) -- (1,0);
  \end{tikzpicture}%
}

 \bibpunct[, ]{(}{)}{,}{a}{}{,}%

\newcommand{\htnp}{\hat{\theta}_{NP}}
\newcommand{\gpe}{\theta^*}

\newcommand{\iid}{\overset{\mathrm{iid}}{\sim}}
\newtheorem{theorem}{Theorem}

\newcommand{\ackname}{Acknowledgements}

\makeatletter
\if@titlepage
  \newenvironment{acknowledgement}{%
    \titlepage
    \null\vfil
    \@beginparpenalty\@lowpenalty
    \begin{center}%
      \bfseries \ackname
      \@endparpenalty\@M
    \end{center}}%
  {\par\vfil\null\endtitlepage}
\else
  \newenvironment{acknowledgement}{%
    \if@twocolumn
      \section*{\ackname}%
    \else
      \small
      \begin{center}%
        {\bfseries \ackname\vspace{-0.5em}\vspace{\z@}}%
      \end{center}%
      \quotation
    \fi}
    {\if@twocolumn\else\endquotation\fi}
\fi
\makeatother

\begin{document}
%%%%%%%%%%%%%%%%
\onehalfspacing

\title{Neutral Pivoting: Strong Bias Correction for Shared Information}

\author{Joseph Rilling\\
	Department of Statistics, Operations, and Data Science\\
	Temple University\\
	Philadelphia, PA  19122, \\
	\texttt{joseph.rilling@temple.edu} \\
 }
	
% Uncomment to remove the date
\date{}

% Uncomment to override  the `A preprint' in the header
\renewcommand{\headeright}{Rilling}
\renewcommand{\undertitle}{}
\renewcommand{\shorttitle}{Neutral Pivoting}

%%% Add PDF metadata to help others organize their library
%%% Once the PDF is generated, you can check the metadata with
%%% $ pdfinfo template.pdf
\hypersetup{
pdftitle={Neutral Pivoting: Strong Bias Correction for Shared Information},
pdfauthor={Joseph Rilling}}

\maketitle
\begin{abstract}
    In the absence of historical data for use as forecasting inputs, decision makers often ask a panel of judges to predict the outcome of interest, leveraging the wisdom of the crowd \citep{woc}. Even if the crowd is large and skilled, shared information can bias the simple mean of judges’ estimates. Addressing the issue of bias, \cite{mp} introduces a novel approach called pivoting. Pivoting can take several forms, most notably the powerful and reliable minimal pivot. We build on the intuition of the minimal pivot and propose a more aggressive bias correction known as the neutral pivot. The neutral pivot achieves the largest bias correction of its class that both avoids the need to directly estimate crowd composition or skill and maintains a smaller expected squared error than the simple mean for all considered settings. Empirical assessments on real datasets confirm the effectiveness of the neutral pivot compared to current methods.
\end{abstract}

\section{Introduction}

Forecasting commonly relies on historical data for predicting future outcomes. In situations where historical data are not readily available/applicable, decision makers may turn to a panel of judges, harnessing the wisdom of the crowd \citep{woc}. By aggregating forecasts from a group of judges, the decision maker hopes to combine the judges’ private informations and produce a more accurate final forecast. However, correlated errors caused by shared information among the judges can cause shared information bias, which does not diminish as the size of the crowd increases. Decision makers rely on accurate estimates to guide their strategies. Therefore it is important that we develop methods to remove shared information bias from aggregated forecasts. We consider these forecasts to be one-off games with no historical information about judge accuracy, and no future opportunity to learn from past mistakes. We must make sure that our method both removes as much bias as possible and limits expected error compared to naïve approaches. 
    
This paper introduces the neutral pivoting estimate. Following the data elicitation first presented in \cite{mp}, and adopted by \cite{so} and \cite{kw}, we ask each judge $j= 1,...J$ to provide a prediction $f_j$ for the mean $\theta$ of the random variable that describes our target of interest. We also ask the judges to provide estimates of the average prediction of their peers, $g_j = [\sum_{k\neq j}f_k]/(J-1)$. The simple means of the individual and peer predictions are given by $\Bar{f}$ and $\Bar{g}$, respectively. The neutral pivot has a concise formulation $\htnp = 3\Bar{f} -2\Bar{g}$, that aggressively removes shared information bias while controlling error in worst case scenarios. The proposed method works for a wide range of crowds, which can differ in both the proportion of the crowd that has private information and the strength of this extra knowledge.  The neutral pivot achieves the largest bias correction of its class that both avoids the need to directly estimate crowd composition or skill, and maintains a smaller expected squared error than the simple mean $\Bar{f}$ for all considered settings.

The structure of this paper is as follows. In Section 2, we discuss the framework as well as alternative methods. In Section 3, we present the neutral pivot and introduce a theorem proving that $\theta_{NP}$ is the largest correction of its type that still guarantees improved performance over the simple mean. We also argue that the bigger correction in $\theta_{NP}$ is a reasonable extension and should be the default bias correction method in its class. We test our method against current approaches in the literature on several experimental datasets in Section 4, and offer final remarks in the conclusion. 

\section{Model Setting and Literature Review}
Our data generating process is the same as the ``nested symmetric" setting from \cite{mp}. In this setting, the decision maker wants to use a crowd of judges to estimate the mean $\theta$ of a random variable $X$. This crowd consists of $ J $ judges, and the judges are divided into two classes: laypeople and mavens. There are $0\leq J_L \leq J$ laypeople and $J_M = J- J_L$ mavens. The proportion of the crowd who are mavens is $p = J_M / J $. Consistent with the literature \citep{mp,so,kw} this proportion is known to the judges but unknown to the decision maker. 

The judges share a common prior distribution $\pi_0$ about $\theta$, characterized by a mean expectation $\mu_0$ and finite variance $\sigma_0$. Depending on their class, a judge has access to either one or two additional signals that contain information about $\theta$. The signals are the sample mean of a certain number of independent and identically distributed observations of $X$. For example, all judges have access the first signal $s_1$, which is called the public signal, and is the mean of $m_1$ observations of $X$. A judge who is a maven receives an additional private signal $t_j$ that is the mean of $l$ private observations of $X$. While the private signals differ among mavens, the sample size $l$ is constant, implicitly assuming that the mavens are all of equal skill. The prior can be related to the signals by assuming that $\mu_0$ is the mean of $m_0$ observations of $X$. With this connection, the shared information $s$, incorporating all information that every judge can access, is the weighted average of the prior mean $\mu_0$ and the public signal $s_1$: $s = (m_0\mu_0 + m_1s_1)/m$, where $m = m_0 + m_1$. All judges know their signals and the sizes of the samples that created them. The decision maker does not know this information. 

The decision maker asks all judges to provide $f_j$, their estimate of $\theta$, and $g_j$, their estimate of the average prediction of the other $J-1$ judges. An optimal layperson $j$, having access only to shared information, submits $f_j$ and $g_j$ both equal to the shared information $s$. Mavens, with access to both shared and private information, have a more complex task. Their optimal prediction $f_j^*$ is a linear combination of shared information and their private signal, with weights according to sample sizes. This optimal prediction for a maven is formulated as $f_j^* = (1-w)s + wt_j$, where $w = l/(l+m)$. All judges know $p$, and a maven incorporates this information into their peer prediction as follows, where $k \neq j$
\begin{align*}
    g_j^* &= E[f_k|s, t_j, p,w] \\
     &= (1-p)E[f_k|s, t_j,w, k =\mathrm{Layperson}] + pE[f_k|s,w,t_j, k =\mathrm{Maven}] \\
     &= (1-p)s + p\{E[(1-w)s + wt_k|s,w,t_j]\} \\
     &= (1-p)s + p\{(1-w)s + wE[t_k|s,w,t_j]\} \\
     &= (1-p)s + p\{(1-w)s + wE[E[t_k|\theta]|s,w,t_j]\} = (1-p)s + p\{(1-w)s + wE[\theta|s,w,t_j]\} \\
     &= (1-p)s + p\{(1-w)s + w[(1-w)s + wt_j]\}.
\end{align*}

\noindent Therefore, the optimal peer prediction for a maven is $g_j^* =(1-pw^2)s + pw^2t_j$. 

As always, deviations from optimal predictions are inevitable. In reality, the judges do not know $p$ or $w$, and cannot be expected to perfectly separate their information into private and public signals. We denote the observed set of predictions by removing the asterisk. With the set of observed predictions $\{f_j, g_j: j=1,...J\}$ the decision maker wants to estimate $\theta$. An optimal estimate for $\theta$ is the global posterior expectation $\gpe = E[\theta|s, t_1,...t_{K},p,w]$, which is the expectation of $\theta$ if the decision maker knew the crowd's compositions, the mavens' skill level, and had access to all the signals individually. If the proportion of mavens in the crowd and the weight of private information are known, \cite{mp} show that, for large crowds, the global posterior expectation is best approximated by

\begin{equation}
    \hat{\theta}^* = \Bar{f} + [1/(pw)](\Bar{f} - \Bar{g}). 
    \label{eqn:gpe}
\end{equation}

However, lacking knowledge of $p$ and $w$, the decision maker faces the challenge of estimating these values, incurring estimation errors. These errors, particularly underestimation of $p$ or $w$, can lead to significant inaccuracies, emphasizing the importance of robust estimation strategies in this complex decision-making process. Therefore, alternative approaches must be designed to approximate $\hat{\theta}^*$ without directly estimating $p$ and $w$. 

We consider three competing models: minimal pivoting \citep{mp}, knowledge weighting \citep{kw}, and the surprising overshoot procedure \citep{so}. Minimal pivoting addresses concerns about estimating $p$ and $w$. The minimal pivot $\hat{\theta}_{MP} = \Bar{f} + (\Bar{f} - \Bar{g})$ makes the smallest possible pivot away from the sample mean. It is equivalent to Equation \ref{eqn:gpe} assuming both $p=1$ and $w=1$, where all judges are mavens and there is no shared information. This approach corrects some bias while avoiding unnecessary risk or adding to the burden on the decision maker.

Knowledge weighting takes a different approach than pivoting. The knowledge weighted estimator does not add or subtract some multiple of $\Bar{f}-\Bar{g}$ to the simple mean. Instead, the knowledge weighted estimator constructs a weighted average of the judge predictions $\hat{\theta}_{KW} = \sum_j \hat{\alpha}_jf_j$, where the $\hat{\alpha}_j$ are designed, in the absence of noise from judges suboptimally reporting $(f_j, g_j)$, to make $\hat{\theta}_{KW}$ equal to ${\theta}^*$ the global posterior expectation.  By individually weighting each judge, the knowledge weighted estimator can take advantage of the varying skill levels and influence of the judges, and adjust for large bias, but also has more challenging estimation than the minimal pivot. 

The surprising overshoot method corrects for bias by using the empirical cumulative distribution and quantile functions of the individual and peer predictions. Its driving theoretical finding is presented as Theorem 4 in \cite{so}: If there exists $f_j \in \{f_1,...,f_J\}$, such that $f_j = \theta$, then $Q(1-p_g) = \theta$, where $Q$ is the quantile function on the individual predictions $f_1,...,f_J$, and  $p_g = \underset{J \rightarrow \infty}{\mathrm{lim}}{\sum_j\mathds{1}\{g_j > \Bar{f}\})}/J $. This approach identifies a relationship between the target of interest and the frequency that the peer predictions overshoot the true average mean. In application, the limit in $p_g$ and the quantile function are both replaced by empirical estimates: $\hat{p}_g = {\sum_j\mathds{1}\{g_j > \Bar{f}\}}/J$ and $\hat{Q}_J(q) = \mathrm{inf}\{f_j \in \{f_1,...f_J\}\mid \hat{F}_J(f_j) > q\}$, and the final estimate is $\hat{\theta}_{SO}=\hat{Q}_J(1- \hat{p}_g)$. The surprising overshoot procedure is notable because it identifies a novel theoretical property for shared information, and achieves strong results while only using empirical distribution/quantile functions, limiting estimation error. 

Each of three competing methods is novel and useful in its own right. Minimal pivoting presents a mathematically sound approach whose simple final formula belies powerful bias reduction. The knowledge weighted estimate introduces a clever way to weight each judge individually, overcoming the lack of historical information about judge skills. Finally, the surprising overshoot algorithm identifies a powerful theoretical property for the probabilistic judgement setting. It is a novel method that avoids introducing undue estimation error. 

\section{Neutral Pivoting}
\label{sec:NP}
Approaches that estimate parameters describing the skill and composition of the crowd have increased volatility, and while these methods have nice theoretical properties, incorrect estimates can lead to poor results in application. The minimal pivot provides stability but undercorrects bias. Neutral pivoting fills the gap in the literature between the minimal pivot and more complex methods, such as pivots that attempt to model the crowd and the knowledge weighted estimate.

Assume without loss of generality that the public information creates a biased simple mean that underpredicts $\theta$. The minimal pivot $\hat{\theta}_{MP} = 2\Bar{f}  - \Bar{g}$ can be used to directly estimate $\theta$ as originally proposed. In truth, the minimal pivot corrects the smallest bias possible according to Equation \ref{eqn:gpe} and estimates a lower bound for the global posterior expectation $\theta^*$. Consider $\hat{\theta}_{MP}^+ = \hat{\theta}_{MP} + \epsilon$, where $\epsilon>0$ is very small. Because minimal pivoting is conservative, $\hat{\theta}_{MP}^+ $ has $\epsilon$ less error than $\hat{\theta}_{MP}$ in almost all cases. The minimal pivot $\hat{\theta}_{MP}$ is only better when $p=w=1$, which corresponds to the unlikely situation where all judges are mavens, and there is no shared information. Even in this rare case,  $\hat{\theta}_{MP}^+$ has only $\epsilon$ more error than minimal pivoting. In other words, if the minimal pivot acts as predicted in \cite{mp} and sets a true lower boundary for $\theta^*$, then there exists a more aggressive estimator $\hat{\theta}_{MP}^+$ that almost completely dominates $\hat{\theta}_{MP}$.

Neutral pivoting builds on this intuition. The minimal pivot gives a lower bound, and estimates that correct for bias more aggressively are likely to be more accurate. The question now becomes ``How much stronger should the estimator be than minimal pivoting?" If the new estimate is not aggressive enough, then it leaves accuracy on the table and only slightly improves the minimal pivot. If the new estimate is too aggressive, it can overshoot $\theta^*$ and become less accurate than the minimal pivot, or even the simple mean. 

The neutral pivot adopts the principle of ``do no harm." It implements the largest correction possible while still guaranteeing lower expected squared error than the simple mean for any crowd composition or skill level. The simple mean is the default for most businesses, researchers, and government agencies. Therefore, the simple mean is a reasonable baseline estimate, and practitioners likely desire guarantees that any new estimate improves on this baseline. The minimal pivot indicates the global posterior expectation is at least $\Bar{f} - \Bar{g}$ away from the mean. A decision maker can be as aggressive as $\Bar{f} + 2(\Bar{f} - \Bar{g})$ while ensuring the estimate is no worse off than the simple mean. Specifically, if both $p=1$ and $w=1$ the absolute error of both the simple mean and the more aggressive estimate is $\Bar{f} - \Bar{g}$. We call our estimate the neutral pivot $\hat{\theta}_{NP} = 3\Bar{f}  - 2\Bar{g}$, because even in the ``worst case" scenario of $p=w=1$, a decision maker should be neutral between $\theta_{NP}$ and $\Bar{f}$. In all other cases, the neutral pivot outperforms the simple mean. Theorem \ref{thm:mse} confirms our statements. 
\begin{theorem}
        Consider the class of estimators $\mathcal{D}  = \{\Bar{f} + \psi(\Bar{f}-\Bar{g})\mid \psi \geq 0\}$. Let $D(\psi) = \Bar{f} + \psi(\Bar{f}-\Bar{g})$ be an instance of this class. Then, as crowd size increases $\lim_{J \to \infty} E[(D(\psi) - \theta)^2] \leq \lim_{J \to \infty} E[(\Bar{f} - \theta)^2]$ for all values of $w\in [0,1]$ and $p \in [0,1]$ if and only if $\psi \in [0,2]$. 
        \label{thm:mse}
\end{theorem}

In the considered class, the simple mean corresponds to $\psi = 0$, and minimal pivoting uses $\psi = 1$. Neutral pivoting sets $\psi = 2$ and is therefore the most aggressive member of the considered class that guarantees expected improvements over the simple mean.

\begin{figure}
    \centering
    \includegraphics[width = 0.75\textwidth]{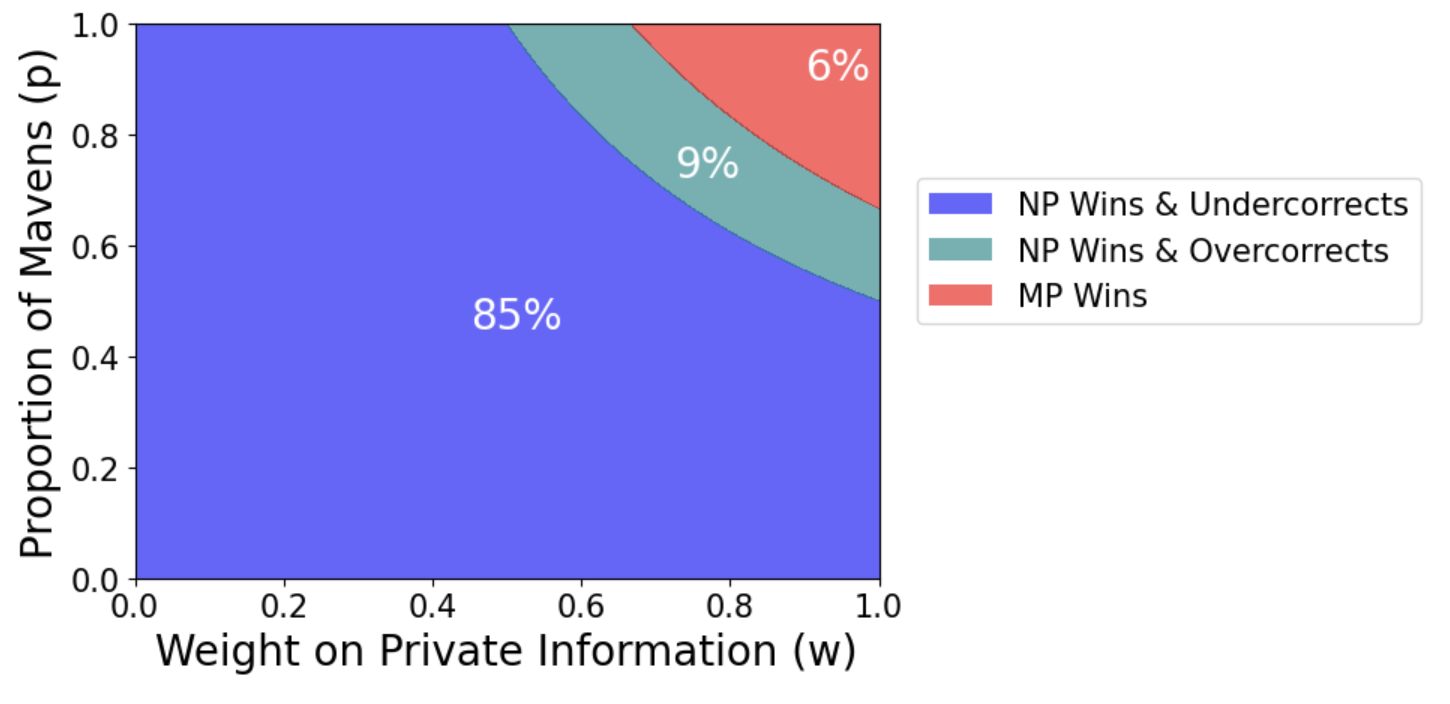}
    \caption{Given a sufficiently large crowd and uninformative priors on crowd composition and skill, the neutral pivot is expected to outperform the minimal pivot $\approx 94\%$ of the time.}
    \label{fig:improv}
\end{figure}

With error controlled, we argue that neutral pivoting should replace minimal pivoting as a default bias correction method for large crowds. Assume the decision maker has no information about the makeup or skill of the crowd. In this case, the decision maker's priors on $p$ and $w$ are independent uniform distributions: $w, p, \iid U[0,1]$. According to Equation \ref{eqn:gpe}, in a large crowd, the neutral pivot is closer than the minimal pivot to the global posterior expectation if $pw \leq 2/3$. With the aforementioned priors, an uninformed decision maker expects the neutral pivot to outperform minimal pivoting with probability $\mathds{P}(pw\leq 2/3) = .937$. Furthermore, neutral pivoting underestimates $\theta^*$ if and only if $pw \leq 1/2$. Therefore, even though the neutral pivot doubles the correction of the minimal pivot, our new estimator is still quite conservative, and the decision maker expects that neutral pivoting under-corrects for bias $\mathds{P}(pw\leq 1/2) = 84.7\%$ of the time. This information is presented graphically in Figure \ref{fig:improv}. The minimal pivot is only more accurate than the neutral pivot in cases where the crowd is both wise (a high proportion of mavens) and has strong private information (high $w$). In cases where there is either moderate shared information or a moderate number of laypeople, the neutral pivot is more accurate because it corrects more for shared information.

\section{Experimental Data}

The following application to data from twelve real world experiments shows neutral pivoting outperforms both naïve aggregation approaches and its most direct competitors: minimal pivoting, knowledge weighting, and the surprising overshoot algorithm. The data for these experiments are made available in the metaggR R package \citep{kw}. The original sources of the data, as well as descriptions of the tasks, are presented in Table \ref{tab:data_description}. The experiments predicting calories and grocery prices have continuous outcomes. The coin flips experiments ask the participants to predict an accurate probability of heads for biased coins. These experiments mimic the symmetric, nested, and nested symmetric information structures proposed in \cite{mp}. The coin experiments allow all judges to view a shared set of coin flip outcomes, and, depending on the desired information structure and role of the judge (layperson/expert/maven), present certain judges with an additional sample of outcomes. The nested symmetric setting is the same as the setting in this paper. The symmetric setting is a special case of the nested symmetric setup, where all judges are mavens ($p=1$). The nested setting is particularly challenging because judges with private information all share the same insight. That is, where mavens receive unique signals $t_j$, the nested case replaces mavens with ``experts" who all share the same private signal $t$. In the nested coin experiment, this constant private information is simulated by giving all experts access to the same additional set of coin flip outcomes. The experiments for NCAA Basketball and General Knowledge have binary outcomes. General knowledge questions are split up into five levels of difficulty, and the basketball dataset is separated by round of the NCAA tournament. 

\begin{table}[h]
\centering
\small
\begin{tabular}{lllll}
\rowcolor[HTML]{4472C4} 
{\color[HTML]{FFFFFF} \textbf{Experiment Class}} & {\color[HTML]{FFFFFF} \textbf{Description}} & {\color[HTML]{FFFFFF} \textbf{\#Judges}} & {\color[HTML]{FFFFFF} \textbf{\#Trials}} & {\color[HTML]{FFFFFF} \textbf{Source}} \\
1. Calorie Counts                                & Calories in pictured food                   & 68-107                                     & 36                                        & \citealp{kw}                                     \\
2. Grocery Prices                                & Price of pictured groceries                 & 49                                         & 10                                        & \citealp{mp}                                     \\
3. Coin Flips (S/N/NS)                           & Probability of heads for coins              & 46-125                                     & 72/24/24                                  & \citealp{mp}                               \\
4. NCAA Basketball (16/64)                       & Winner of tournament games                  & 48-165                                     & 24/96                                     & \citealp{mart}                                  \\
5. General Knowledge (5×D)                       & Binary choice trivia                        & 89-95                                      & 100×5                                     & \citealp{mart}                              
\end{tabular}
\normalsize
\caption{There are twelve experiments split into five classes. The coin flips class has three subtypes: symmetric, nested, and nested symmetric. NCAA basketball predictions are for the round of 16 and separately for the round of 64. General knowledge is split into five categories by difficulty, with 100 questions in each difficulty level.  }
\label{tab:data_description}
\end{table} 

The root mean squared errors (RMSE) of each method for each experiment are presented in Table \ref{tab:results}. The table also includes results for the trimmed average \citep{hillier_combining_2001, stock_combination_2004, jose_simple_2008}. In this specific case, the trimmed average is the mean of $f_j$, $j=1,..,J$ when the upper and lower 10\% of submissions are removed \citep{kw}. For probabilistic/binary prediction tasks, method outputs are Winsorized to valid answers in $[0,1]$. Neutral pivoting outperforms all competitors in eight of the twelve experiments. The procedure beats the simple mean in all experiments, and does so significantly in all but two experiments. Even in cases with less private information, such as the nested setting where Theorem \ref{thm:mse} does not apply, the procedure performs comparably to minimal pivoting and knowledge weighting.

\begin{table}[h]
    \centering
    \begin{tabular}{c|c|c|c|c|c|c|c}
        Study & Mean & Median & Trimmed & Minimal & Knowledge &Surprising& Neutral  \\ 
        &&&Average&Pivoting&Weighted& Overshoot &Pivoting  \\ \hline
        Calorie Counts&     419.329 &   436.532    &  443.693 & 399.054*    & 393.841*+  & 402.806* &\underline{\textit{381.927}}*+ \\
        Grocery Prices&     8.901&      9.134        &  8.992 &  8.322*   &  8.171*             & 8.290& \underline{\textit{7.788}}*+\\
        Coin Flips, Sym&    7.526 &     7.842       &  7.684 &  5.339*  &  5.281*               & \underline{\textit{5.279}}*&5.286*\\
        Coins Flips, N&     9.295&      10.270       &  9.402 & 8.532   &  8.674    &\underline{\textit{8.088}}           & 8.543\\
        Coin Flips, NS&     12.770 &    14.532         &  13.592 & 10.737*  &  10.032*+      &9.922*    & \underline{\textit{9.361}}*+\\
        NCAA R16&           .436&       .453                  &  .436 & .429  &  .424         & .438&\underline{\textit{.423}}\\
        NCAA R64&           .434&       .434                 &  .432 & .431*  &  .430*+ & \underline{\textit{.428}} &.431*+\\
        GK1&                .319 &      .286 &  .298    & .274*     &  .260*+     &      \underline{\textit{.236}}*+          &.238*+\\
        GK2&                .385 &      .390 &  .378    & .343*    & .326*+          & .310*+          & \underline{\textit{.308}}*+\\
        GK3&                .428&       .435 &  .427    & .394*    &  .377*+  & .373*+                  &\underline{\textit{.364}}*+\\
        GK4&                .459&       .467 &  .460    &  .439*    &  .430*+   &.433*               &\underline{\textit{.423}}*+\\
        GK5&                .460&       .466 &  .462    &  .440*    &  .431*+    & .440*               &\underline{\textit{.422}}*+\\
    \end{tabular}
    \caption{Underline and italics indicates best on the table. * indicates p-value $<0.1$ for a Wilcoxon signed rank test of simple mean vs MP, KW, SO, NP. + indicates p-value $<0.1$ for a Wilcoxon signed rank test MP vs KW, SO, NP.}
    \label{tab:results}
\end{table}

For each dataset, we also calculate $\psi_O$, the oracle $\psi$ that gives the lowest mean squared error. The oracle can be calculated using the true value $\theta$ according to $\psi_O = \sum_E\big[(\Bar{f}- \Bar{g})(\theta - \Bar{f})\big]/(\sum_E{(\Bar{f}- \Bar{g})^2})$, where $E$ is the set of experiments in the dataset. The oracle values and resulting RMSE are presented in Table \ref{tab:oracle}. The calculated $\psi_O$ is always greater than one, supporting the idea that the minimal pivot gives a conservative bound for the bias correction. Furthermore, nine of the twelve experiments have an oracle $\psi_O$ greater than two, and half of the considered experiments have an oracle $\psi_O$ greater than five. This result supports the idea that while the neutral pivot doubles the correction of the minimal pivot, it is often still conservative and undercorrects for bias. While these findings are preliminary, they lend some support to the argument described in Section \ref{sec:NP} and summarized in Figure \ref{fig:improv}. In the future, a broad experimental study that calculates oracle pivots across a wider variety of settings and crowds would contribute to the literature on pivoting.

\begin{table}[h]
    \centering
    \begin{tabular}{|c|c|c|}
        \hline
        \textbf{Experiment} & $\psi_O$ & RMSE \\
        \hline
        Calorie Counts & 5.657 & 352.636 \\
        Grocery Prices & 7.795 & 6.141\\
        Coins Sym & 1.520 & 4.978\\
        Coins N & 1.486 & 8.436\\
        \hline
    \end{tabular}
    \quad
\begin{tabular}{|c|c|c|}
        \hline
        \textbf{Experiment} & $\psi_O$ & RMSE \\
        \hline
        Coins NS & 2.873 & 8.941 \\
        NCAA16 & 5.692 & .414\\
        NCAA64 & 1.634 & .431\\
        GK1 & 3.545 & .216 \\
        \hline
    \end{tabular}
    \quad
    \begin{tabular}{|c|c|c|}
        \hline
        \textbf{Experiment} & $\psi_O$ & RMSE \\
        \hline
        GK2 & 4.032 & .278 \\
        GK3 & 5.148 & .321\\
        GK4 & 5.374 & .398\\
        GK5 & 5.713 & .393\\
        \hline
    \end{tabular}
    \caption{Oracle Pivots and RMSE for each experiment. None of the optimal pivots are below 1. Nine of the twelve experiments have an oracle $\psi_O$ above 2, indicating that, despite its aggressive correction, the neutral pivot still undercorrects for bias in the majority of the considered datasets.}
    \label{tab:oracle}
\end{table}

\begin{figure}[h]%
    \centering
    \subfloat[\centering Calories \label{fig:rcal}]
    {{\includegraphics[width=0.24\textwidth]{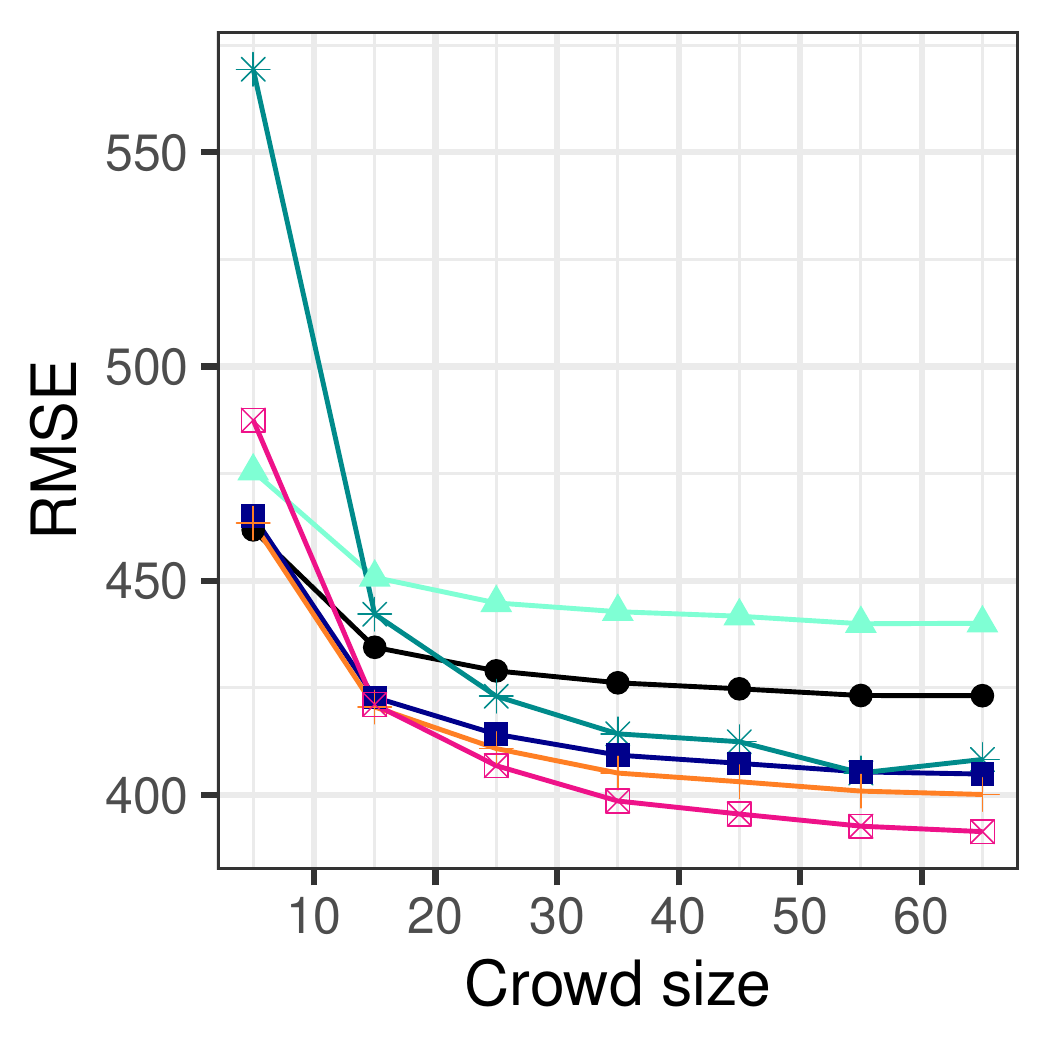} }}%
    \subfloat[\centering Groceries \label{fig:rgroc}]
    {{\includegraphics[width=0.24\textwidth]{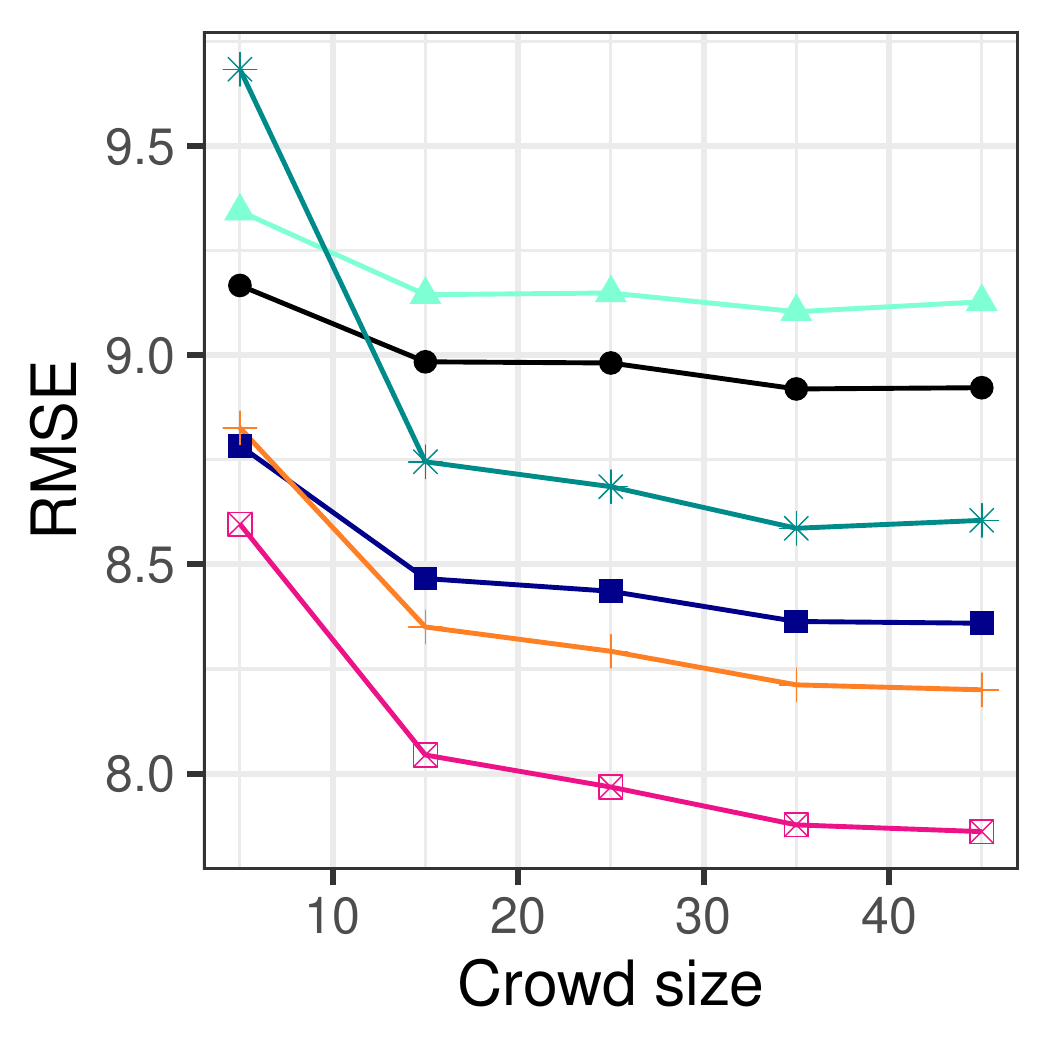} }}%
    \subfloat[\centering GK Diff. 1\label{fig:rgk1}]{{\includegraphics[width=0.24\textwidth]{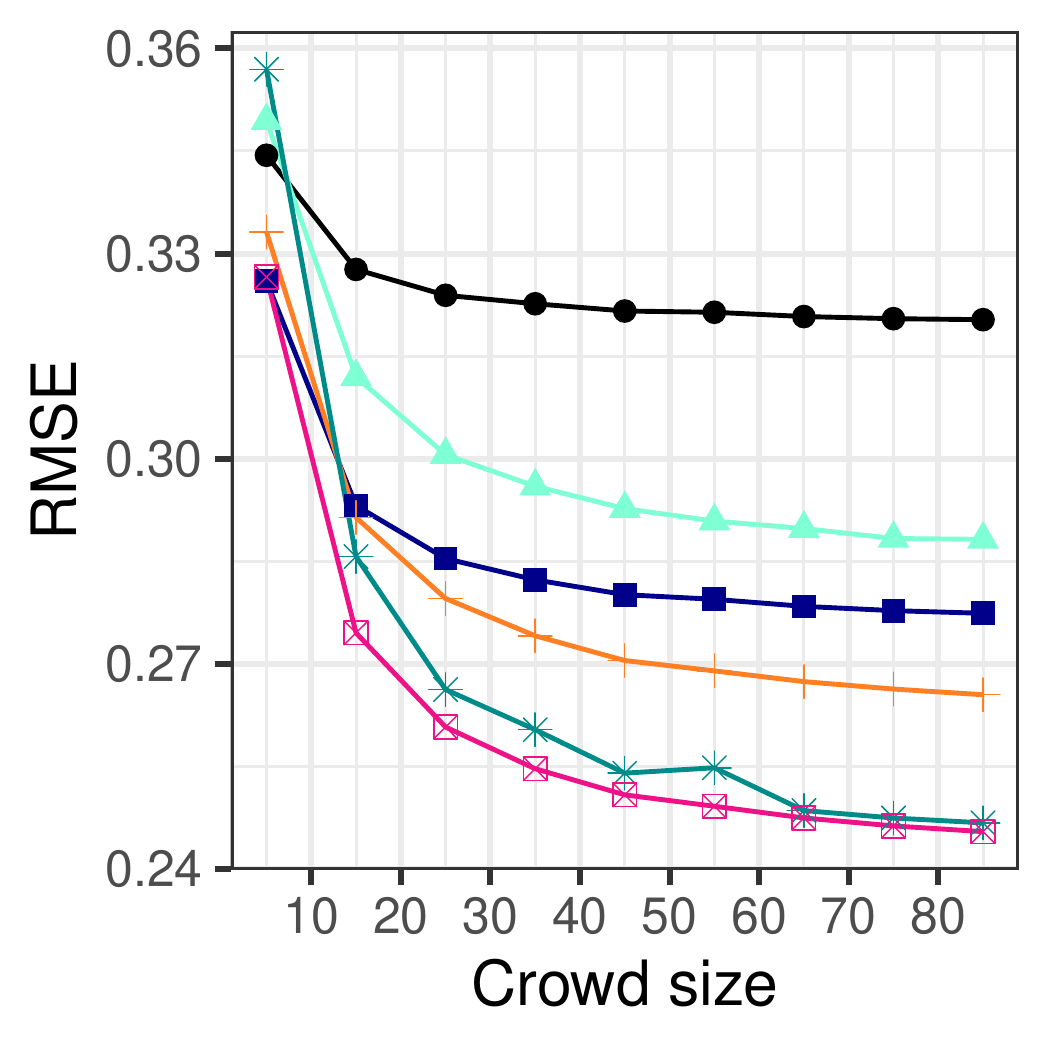} }}%
    \subfloat[\centering GK Diff. 2\label{fig:rgk2}]{{\includegraphics[width=0.24\textwidth]{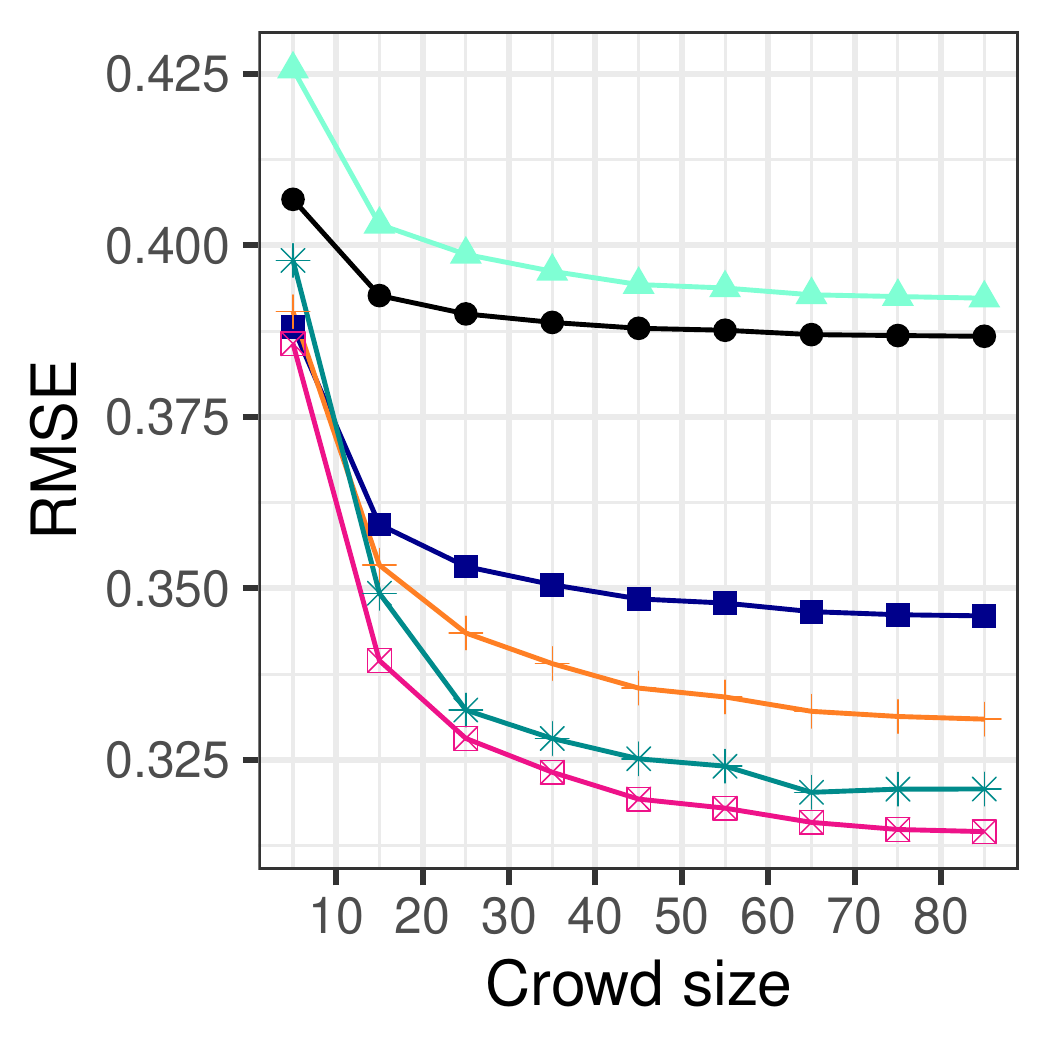} }}%
    \\
    \subfloat[\centering GK Diff. 3\label{fig:rgk3}]{{\includegraphics[width=0.24\textwidth]{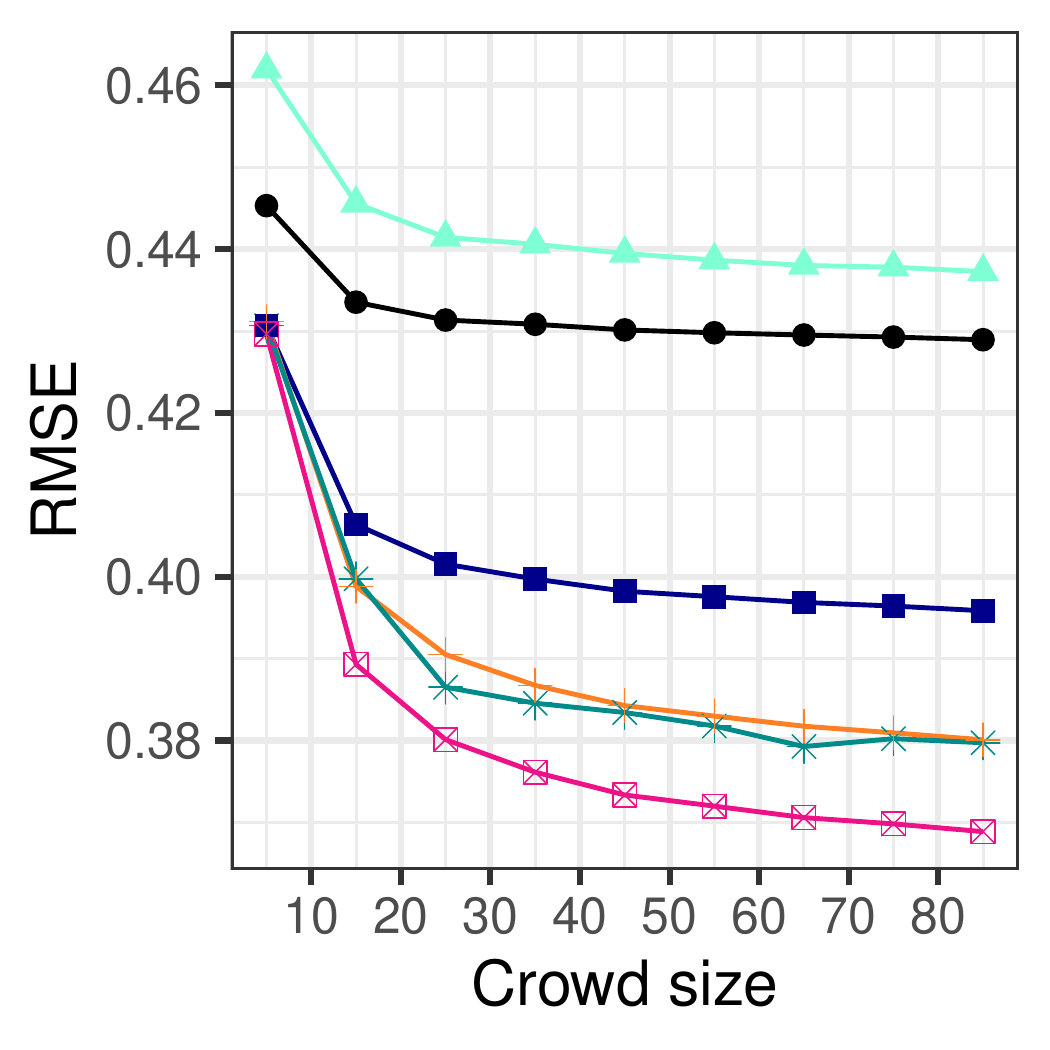} }}
    \subfloat[\centering GK Diff. 4\label{fig:rgk4}]{{\includegraphics[width=0.24\textwidth]{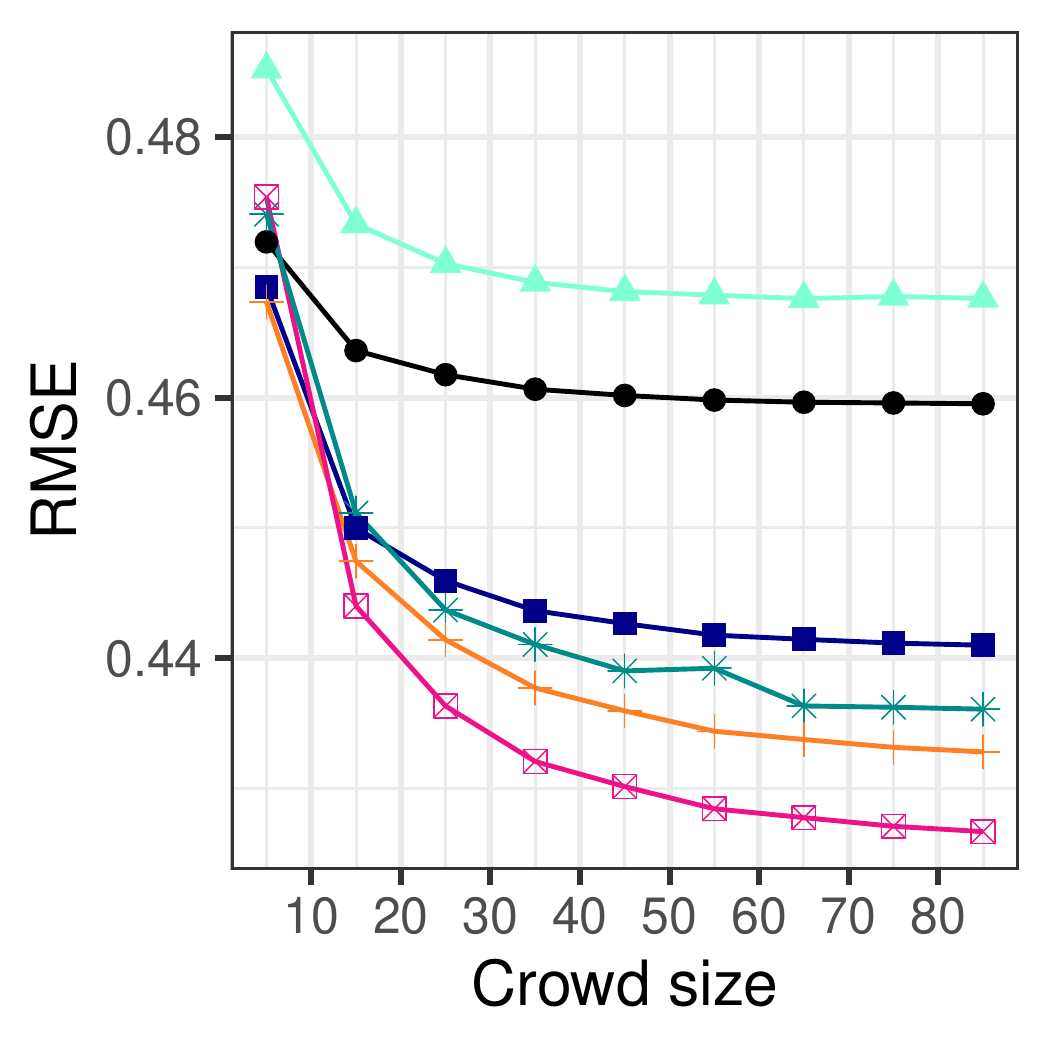} }}%
    \subfloat[\centering GK Diff. 5\label{fig:rgk5}]{{\includegraphics[width=0.24\textwidth]{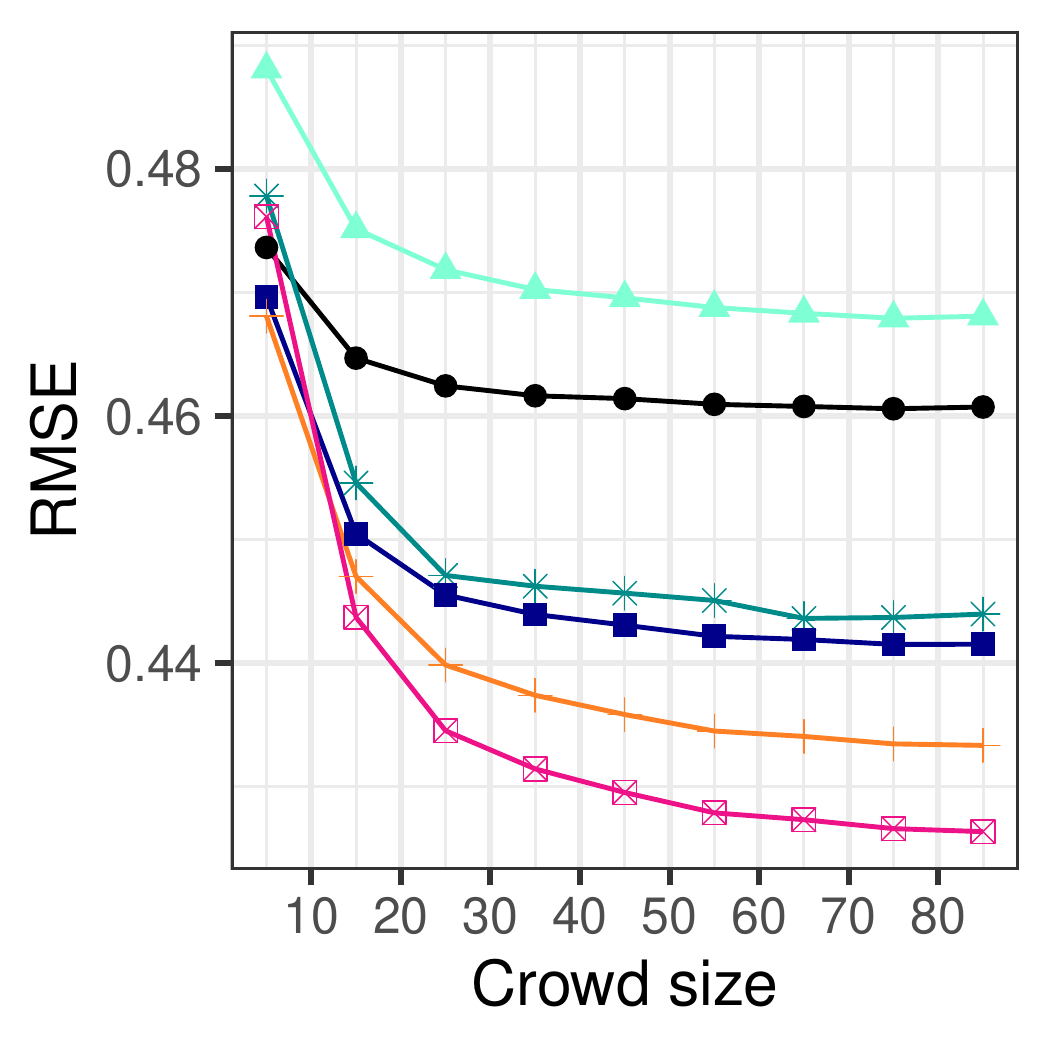} }}%
    \subfloat[\centering NCAA Round of 16 \label{fig:rncaa16}]
    {{\includegraphics[width=0.24\textwidth]{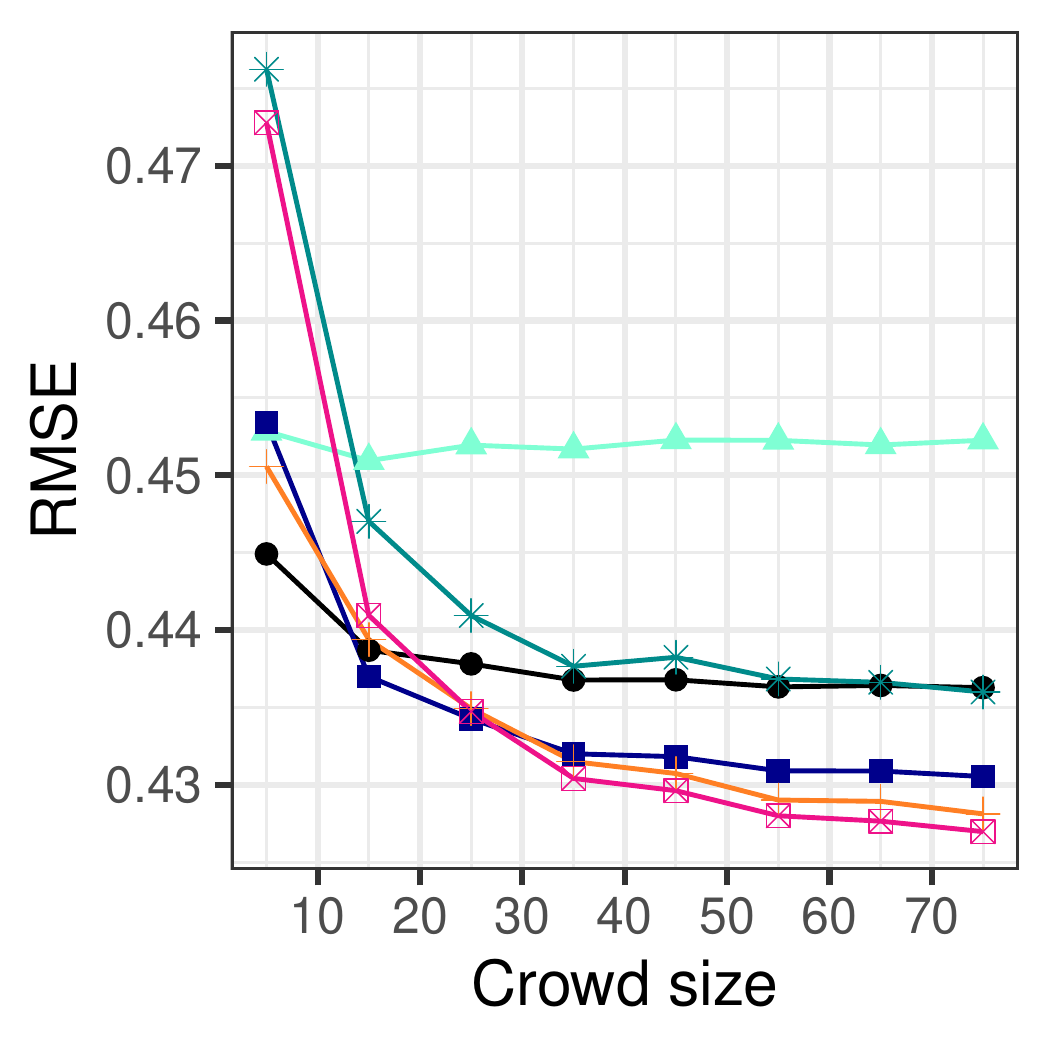} }}%
    \\ 
    \subfloat[\centering NCAA Round of 64 \label{fig:rncaa64}]
    {{\includegraphics[width=0.24\textwidth]{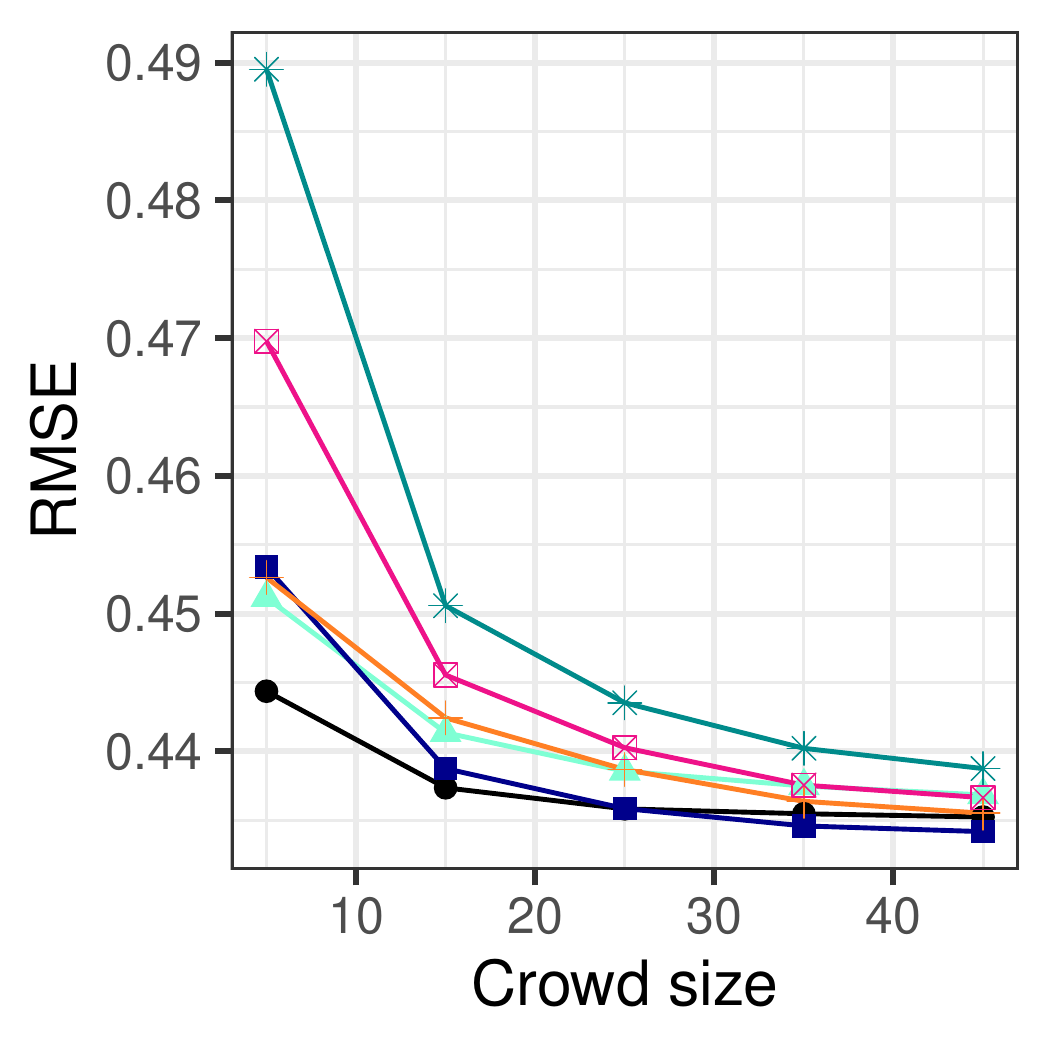} }}%
    \subfloat[\centering Coins Sym. \label{fig:rs}]{{\includegraphics[width=0.24\textwidth]{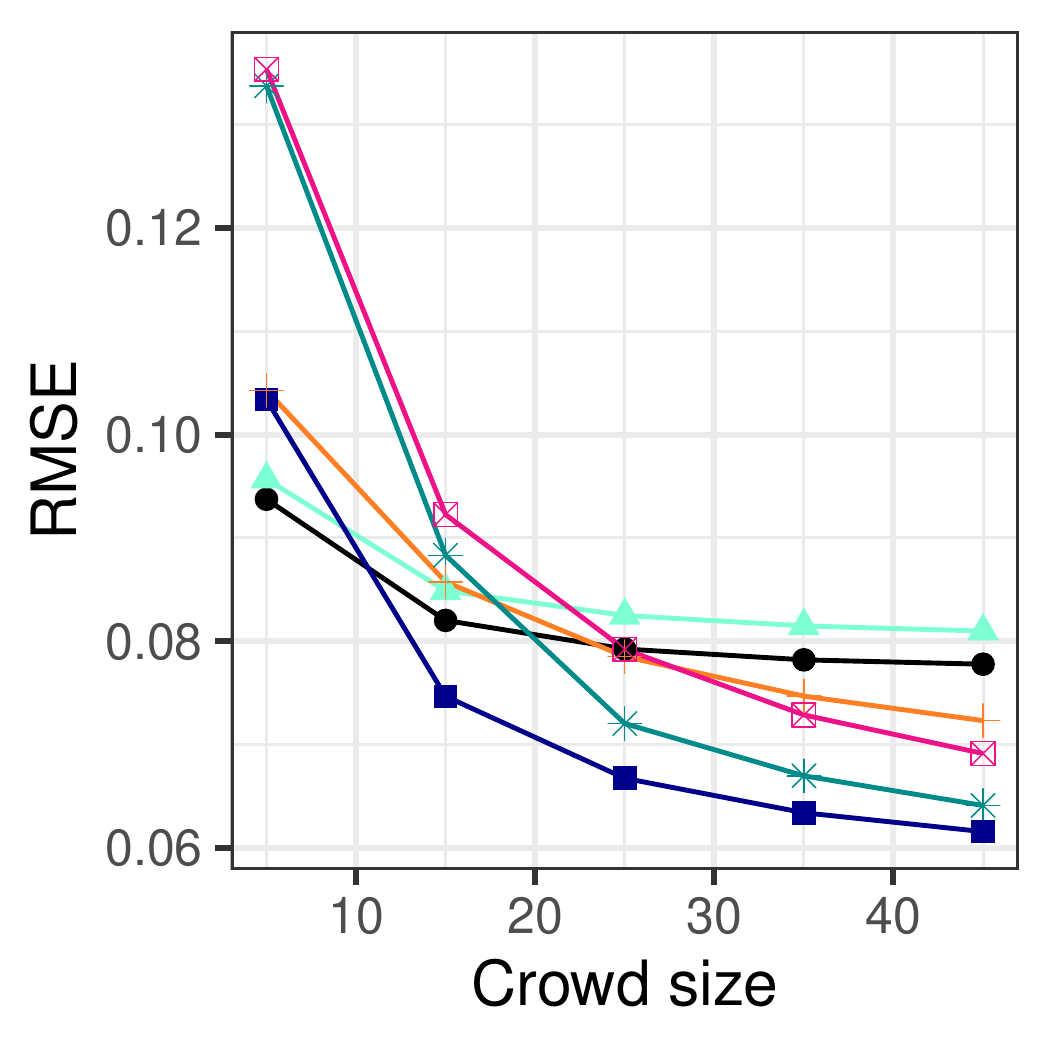} }}%
    \subfloat[\centering Coins Nested Sym. \label{fig:rns}]{{\includegraphics[width=0.24\textwidth]{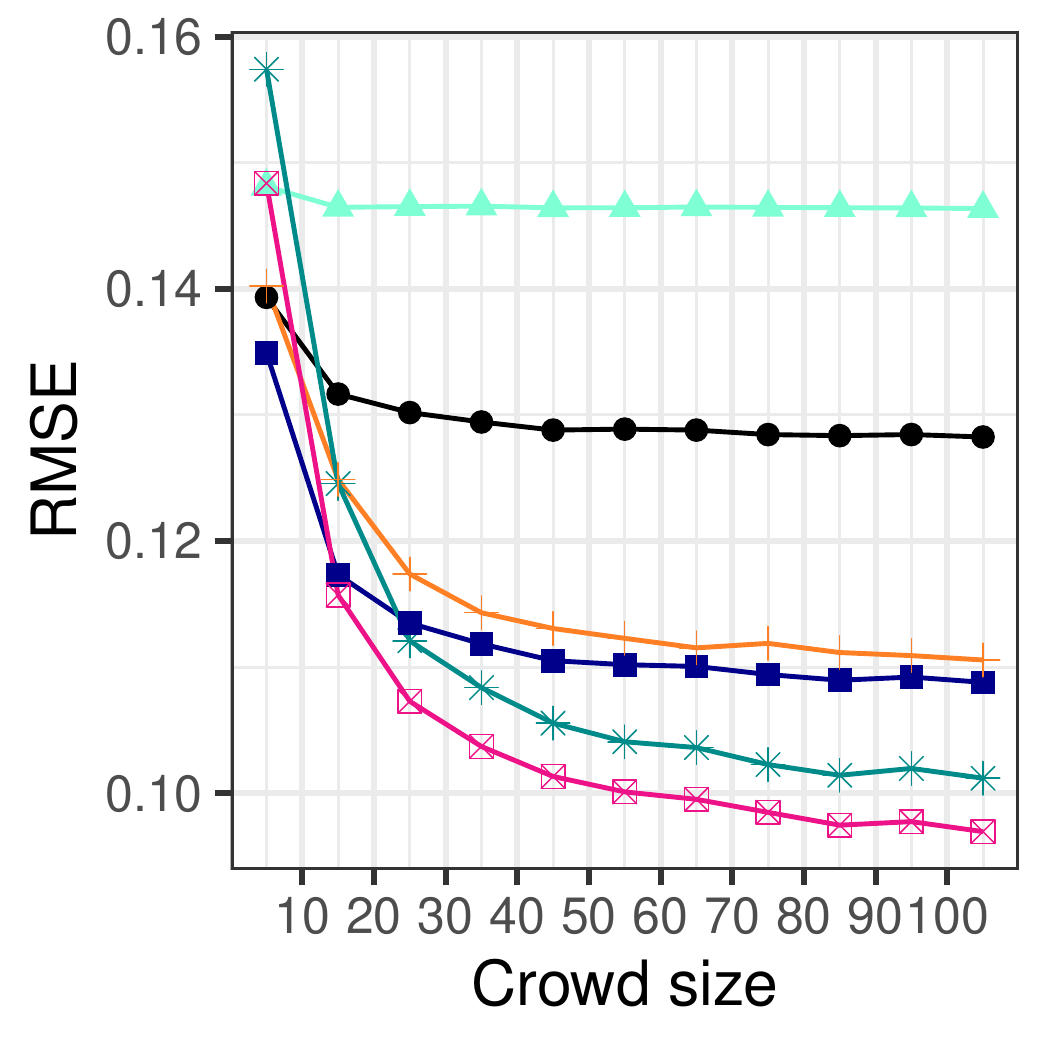} }}%
    \subfloat[\centering Coins Nested \label{fig:rn}]{{\includegraphics[width=0.24\textwidth]{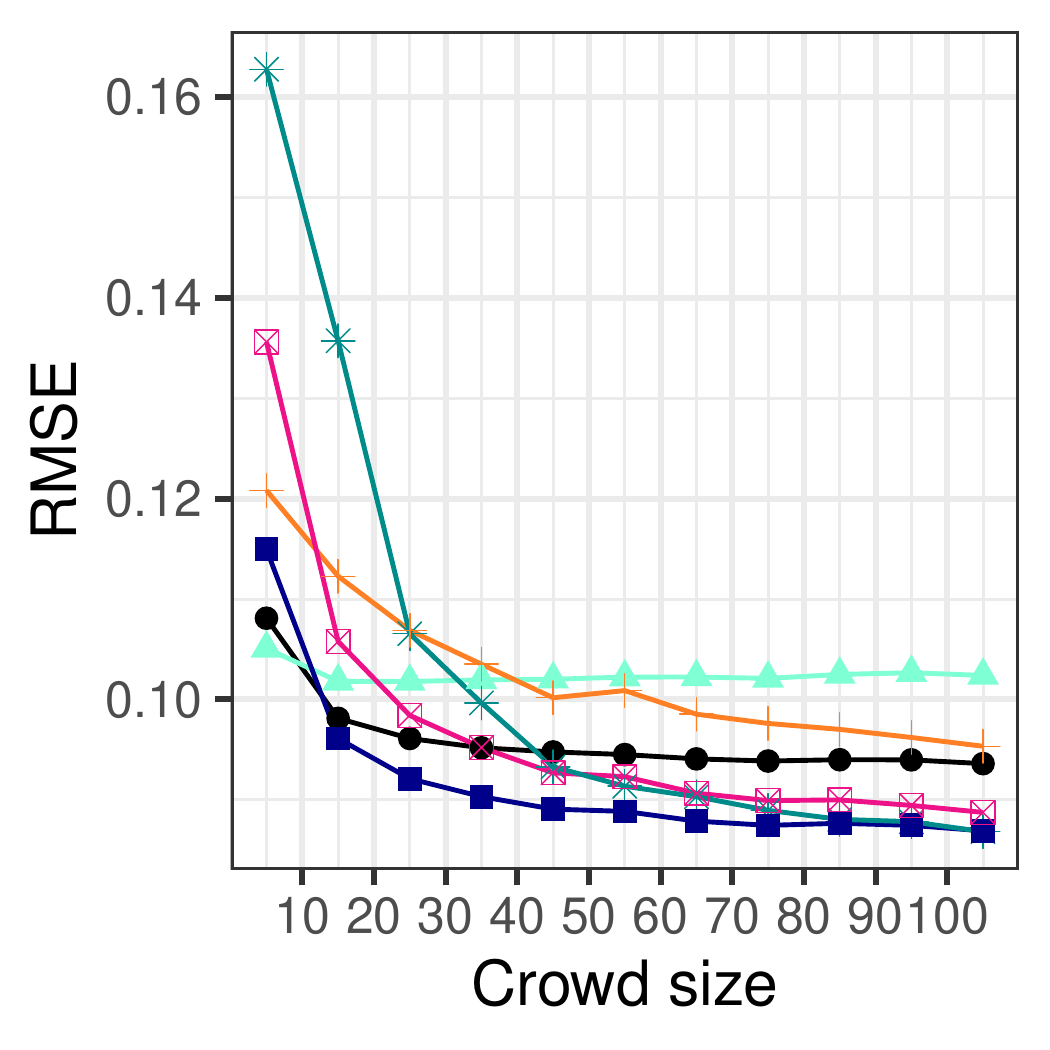} }}%

\caption{Bootstrap root mean squared error. \textcolor{blue}{MP} ($\blacksquare$), \textcolor{YellowOrange}{KW}($+$), \textcolor{magenta}{NP} (\usebox{\mytikzbox}), \textcolor{black}{Simple Average} ($\CIRCLE$), \textcolor{SkyBlue}{Median} ($\blacktriangle$), \textcolor{TealBlue}{SO} ($*$)}
    \label{fig:rbootstrap}%
\end{figure}

Finally, we also examine how methods perform as a function of crowd size. We use bootstrap resampling on each experiment to obtain crowds of different sizes. Figure \ref{fig:rbootstrap} plots each method's average RMSE for one thousand bootstrapped crowds of a given size. Figure \ref{fig:bootstrap} in Appendix \ref{a:rmse_ratio} gives the same information as the ratio of competing methods' RMSE divided by the neutral pivot's RMSE. The plots strengthen the empirical evidence supporting neutral pivoting. The neutral pivot struggles in very small crowds ($J \approx 5$), but strengthens rapidly as crowd size increases. The volatility in small crowds is likely due to sampling errors in the sample means of the individual and peer predictions. If the estimation of these quantities is inaccurate when compared to the true population means, this inaccuracy is passed onto the neutral pivot and can lead to poor results. However, sample means are generally stable and converge rapidly, and moderate crowd sizes quickly minimize uncertainty in the lower bound.  

This application study supports our claims about neutral pivoting. Minimal pivoting offers great bias reduction and strong theoretical guarantees. Neutral pivoting corrects more bias than its predecessor, and also maintains control of error, as shown in Theorem \ref{thm:mse}. The approach remains easy to explain to practitioners, is simple to implement, and avoids estimation error that is common in more complicated models. 

\section{Conclusion}

The neutral pivot is a natural extension of minimal pivoting procedure. It is simple, powerful, and has an intuitive control of error aligned with the notion of ``do no harm" that practitioners may find quite appealing. Shared information bias is present in many fields. This clear and concise correction can significantly improve predictions without the need for historical data. Furthermore, the idea of using minimal estimates as inputs for more aggressive approaches may find broader application in low data environments, where large estimation errors undercut complex approaches. Finally, this work and its peers assume symmetric loss functions. Further research may focus on allowing practitioners to specify their potentially asymmetric loss functions, and estimate an optimal pivot size that minimizes their expected loss.
% Acknowledgments here
\clearpage
\begin{acknowledgement}
    I gratefully acknowledge the help of Alan Karr. His comments and guidance during the editing process have greatly improved this paper. I also thank Ken McAlinn for various discussions concerning ensemble methods and I thank Cem Peker for providing his code as a starting point for the bootstrapped crowd size simulations. Finally, I thank an anonymous AE and two anonymous reviewers for feedback on an earlier version of this work. This research includes calculations carried out on HPC resources supported in part by the National Science Foundation through major research instrumentation grant number 1625061 and by the US Army Research Laboratory under contract number W911NF-16-2-0189.
\end{acknowledgement}

\begin{appendix}
\section{Proof of Theorem 1}
\label{sec:thm1}
We adopt the nested symmetric structure of \cite{mp}. Let all notation be the same as in the paper. As mentioned in the paper, judges report their predictions and peer predictions with some degree of error. For laypeople, there is error $\delta_j$, where $f_j = g_j = s - \delta_j$. Mavens have two types of error. They report their predictions via $f_j = f_j^* + \epsilon_j$, and have additional error $\gamma_j$ in their peer predictions: $g_j = (1-pw^2)s + pw^2t_j + pw\epsilon_j + \gamma_j$. The errors $\{\delta_j, \epsilon_j, \gamma_j\}$ are zero mean, independent for all $j$, and independent between all $j= 1:J$. For simplicity of notation, we call the set of errors $\kappa_j = \{\epsilon_j, \delta_j, \gamma_j\}$. Under this structure, \citeauthor{mp} showed that the simple mean of the individual and peer predictions can be written as follows:
\begin{equation}
    \Bar{f} = (1-pw)s + pw\sum_{j=1}^{pJ}(t_j/(pJ)) + (1-p)\sum_{j=1}^{J-pJ}(\delta_j/(J-pJ))+ p\sum_{j=1}^{pJ}(\epsilon_j/(pJ))
    \label{eq:fbar}
\end{equation}
and 
\begin{equation}
    \Bar{g} = (1-p^2w^2)s + p^2w^2\sum_{j=1}^{pJ}(t_j/(pJ)) +(1-p)\sum_{j=1}^{J-pJ}(\delta_j/(J-pJ))+ p^2w\sum_{j=1}^{pJ}(\epsilon_j/(pJ)) + p\sum_{j=1}^{pJ}(\gamma_j/(pJ) ).
    \label{eq:gbar}
\end{equation}
Consider the class of estimators, $\mathcal{D}  = \{\Bar{f} + \psi(\Bar{f}-\Bar{g})\mid \psi \geq 0\}$. Denote any particular instance with a subscript: $\hat{\theta}_{\psi} = \Bar{f} + \psi(\Bar{f}-\Bar{g})$. Substitute the expanded definitions of $\Bar{f}$ and $\Bar{g}$ from Equations \ref{eq:fbar} and \ref{eq:gbar} respectively to obtain the following
\begin{align*}
    \hat{\theta}_{\psi} &= (1+\psi)\Bar{f} - \psi\Bar{g} \\
    &= (1 -(1+\psi)pw  + \psi p^2w^2)s + ((1+\psi)pw - \psi p^2w^2)\sum_{j=1}^{pJ}(t_j/(pJ)) + (1-p)\sum_{j=1}^{J-pJ}(\delta_j/(J-pJ)) \\
    &\quad \quad + ((1+\psi)p-\psi p^2w)\sum_{j=1}^{pJ}(\epsilon_j/(pJ)) -\psi p\sum_{j=1}^{pJ}(\gamma_j/(pJ) ). 
\end{align*}
The expected mean squared error is $E[(\hat{\theta}_{\psi}- \theta)^2]$. The expectation can be broken into an inner expectation over the signals $s_1, \{t_1,...,t_{pJ}\}$, and the errors $\kappa$, and an outer expectation on $\theta$ with respect to the prior $\pi_0$. This reformulation gives $E[(\hat{\theta}_{\psi} - \theta)^2] = E_{\theta \sim \pi_0}[E_{s_1,t, \kappa}[(\hat{\theta}_{\psi} - \theta)^2|\theta]]$. 
\begin{align*}
    E[(\hat{\theta}_{\psi} - \theta)^2] &= E[E[ \{(1-(1+\psi)pw + \psi p^2w^2)(\frac{m_0}{m}(\mu_0 -\theta) + \frac{m_1}{m}(s_1-\theta)) + \frac{pw((1+\psi) - \psi pw)}{pJ}\sum_{j=1}^{pJ}(t_j - \theta) \\ & \quad \quad + \frac{1}{J}\sum_{j=1}^{J-pJ}(\delta_j-0)  + (((1+\psi)-\psi pw)/J)\sum_{j=1}^{pJ}(\epsilon_j-0) -(\psi /J)\sum_{j=1}^{pJ}(\gamma_j-0)\}^2|\theta]]. 
\end{align*}
Take the inner expectation with respect to $s_1, t,$ and $\kappa$. Because $s_1$ and $t$ are assumed to come from samples of $X$, their expectation is $\theta$. The errors are independent and zero mean. Therefore, all squared differences, except those of the prior mean, become variances
\begin{align*}
    E[(\hat{\theta}_{\psi} - \theta)^2] &= E[(1-(1+\psi)pw + \psi p^2w^2)^2(\frac{m_0^2}{m^2}(\mu_0 - \theta)^2 + \frac{m_1^2}{m^2}Var(s_1|\theta))\\ & \quad \quad + \frac{w^2((1+\psi)-\psi pw)^2}{J^2}\sum_{j=1}^{pJ}Var(t_j|\theta)  + \frac{1}{J^2}\sum_{j=1}^{J-pJ}Var(\delta_j)  \\
    & \quad \quad +\frac{((1+\psi)-\psi pw)^2}{J^2}\sum_{j=1}^{pJ}Var(\epsilon_j) + \frac{\psi ^2}{J^2}\sum_{j=1}^{pJ}Var(\gamma_j)]. 
\end{align*}

Take the expectation over the prior, and use the independent and identically distributed properties of the maven signal and errors to remove the summations
\begin{align*}
    E[(\hat{\theta}_{\psi} - \theta)^2] &= (1-(1+\psi)pw + \psi p^2w^2)^2(\frac{m_0^2\sigma_0^2 + m_1^2(V_0/m_1)}{m^2}) +  \frac{pw^2((1+\psi)-\psi pw)^2}{J}(V_0/l) \\
    &\quad \quad + \frac{1-p}{J}Var(\delta) + \frac{p((1+\psi)-\psi pw)^2}{J}Var(\epsilon) + \frac{\psi ^2p}{J}Var(\delta).
\end{align*}

As $J \rightarrow \infty$, both the coefficient multiplying the variance of the mavens' signals, and the coefficients of the variances of the errors terms go to zero. Therefore, as $J \rightarrow \infty$, $E[(\hat{\theta}_{\psi} - \theta)^2] \rightarrow (1-(1+\psi)pw + \psi p^2w^2)^2(m_0^2\sigma_0^2 + m_1^2(V_0/m_1))/{m^2}$. \cite{mp} prove the large crowd limiting expected squared error of the simple mean is  $E[(\Bar{f} - \theta)^2] \rightarrow (1-pw)^2({m_0^2\sigma_0^2 + m_1^2(V_0/m_1)})/{m^2}$. Therefore, if $(1-pw)^2 \geq (1-(1+\psi)pw + \psi p^2w^2)^2$ for all $p,w \in (0,1)^2$, then the expected square error of $\hat{\theta}_{\psi}$ is less than or equal to the expected squared error of $\Bar{f}$.  The inequality is presented below
\begin{align*}
    (1-pw)^2 \geq (1-(1+\psi)pw + \psi p^2w^2)^2.
\end{align*}
If $pw = 0$, both sides of the inequality are 1 and independent of $\psi$, so all $\psi$ work. If $pw = 1$, then both sides of the inequality are 0 and independent of $\psi$, so all $\psi$ work again. If $pw\in (0,1)$, then the inequality can be rewritten as
\[pw -1 \leq 1-(1+\psi)pw + \psi p^2w^2) \leq 1-pw.\]

Through standard algebra techniques, the inequality simplifies to 
\[-2 \leq -pw\psi  \leq 0.\]
This final line is true for all $p,w \in (0,1)$ if and only if $\psi \in [0,2]$. Given that neutral pivoting corresponds to $\psi = 2$, it is the most aggressive procedure that has lower expected squared error than the sample mean.
\clearpage
\section{RMSE Ratio Plots}
\label{a:rmse_ratio}
\begin{figure}[h]%
    \centering
    \subfloat[\centering Calories \label{fig:cal}]
    {{\includegraphics[width=0.20\textwidth]{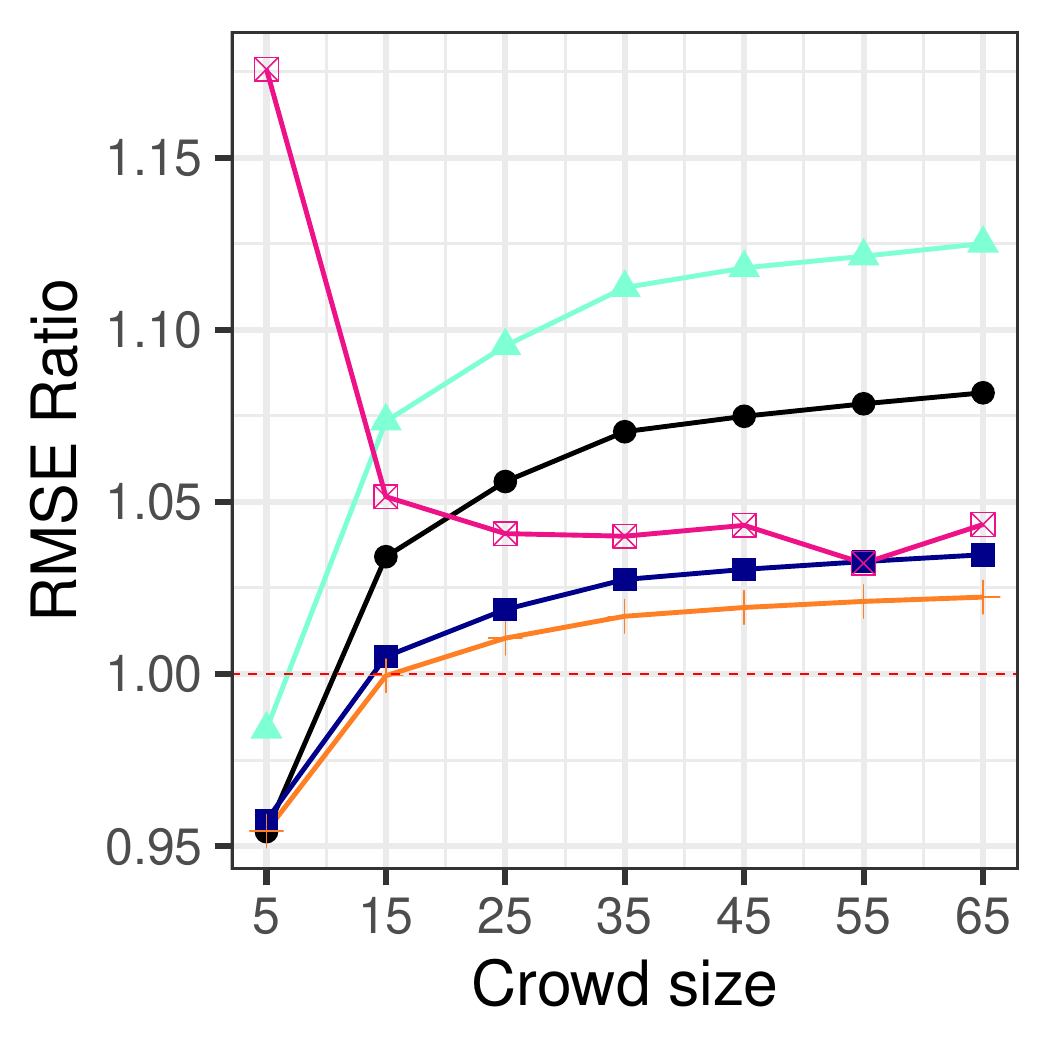} }}%
    \quad
    \subfloat[\centering Groceries \label{fig:groc}]
    {{\includegraphics[width=0.20\textwidth]{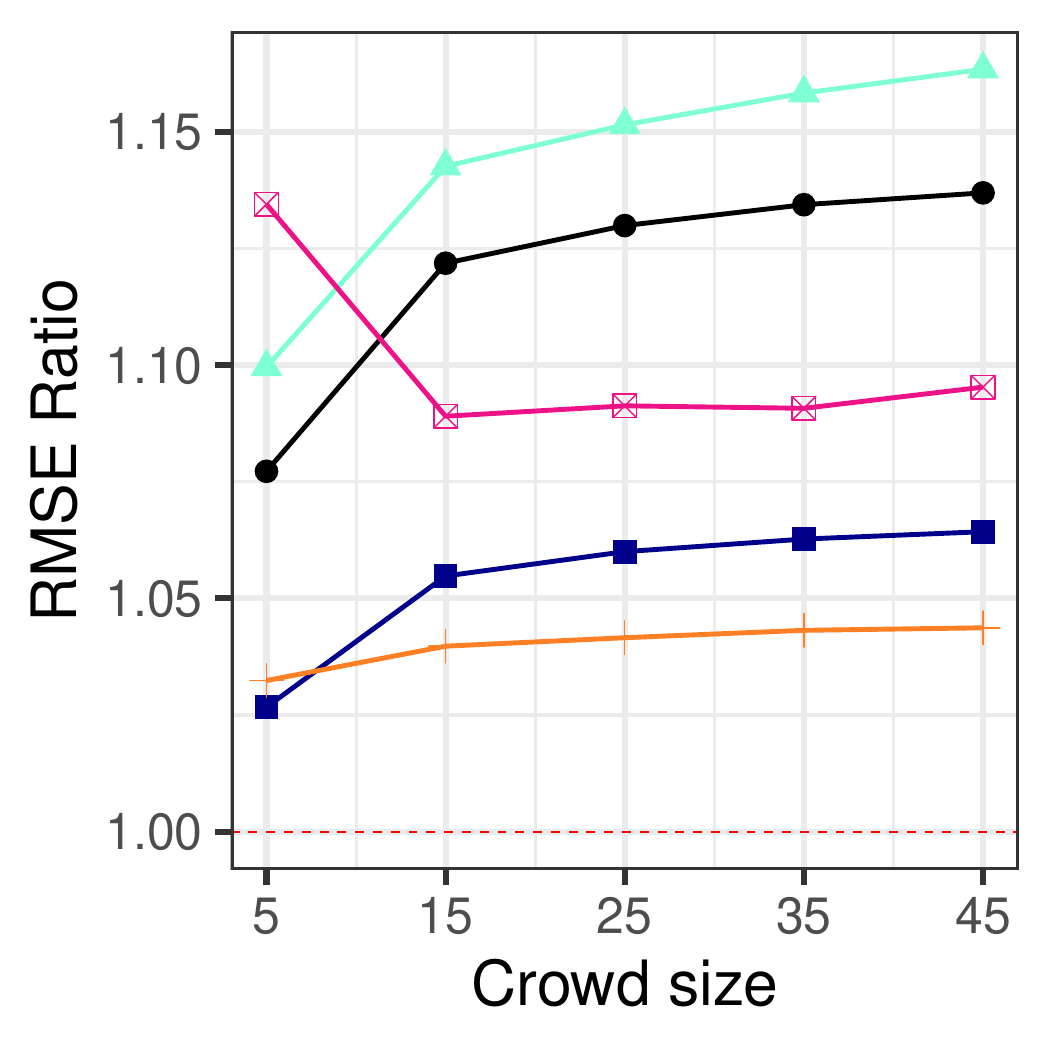} }}%
    \quad 
    \subfloat[\centering GK Diff. 1\label{fig:gk1}]{{\includegraphics[width=0.20\textwidth]{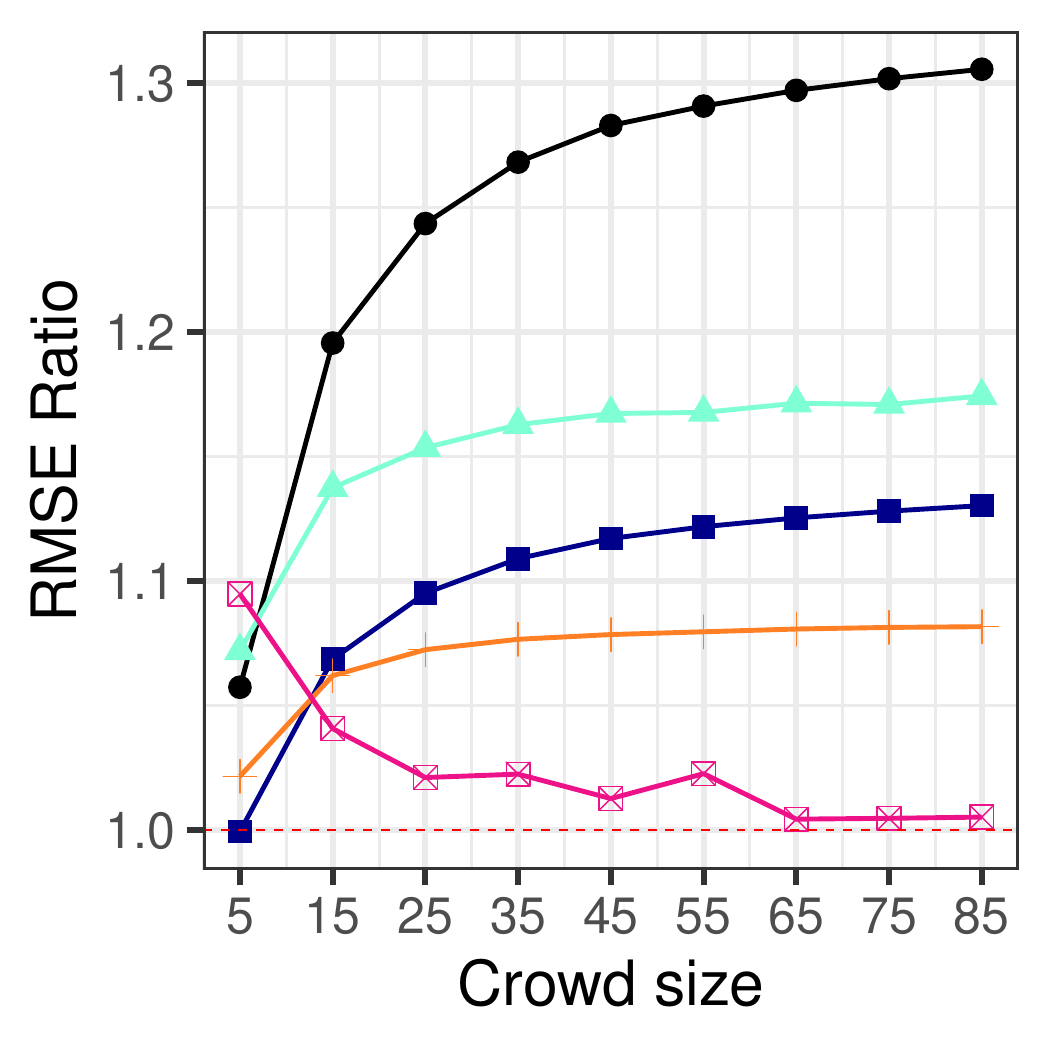} }}%
    \quad
    \subfloat[\centering GK Diff. 2\label{fig:gk2}]{{\includegraphics[width=0.20\textwidth]{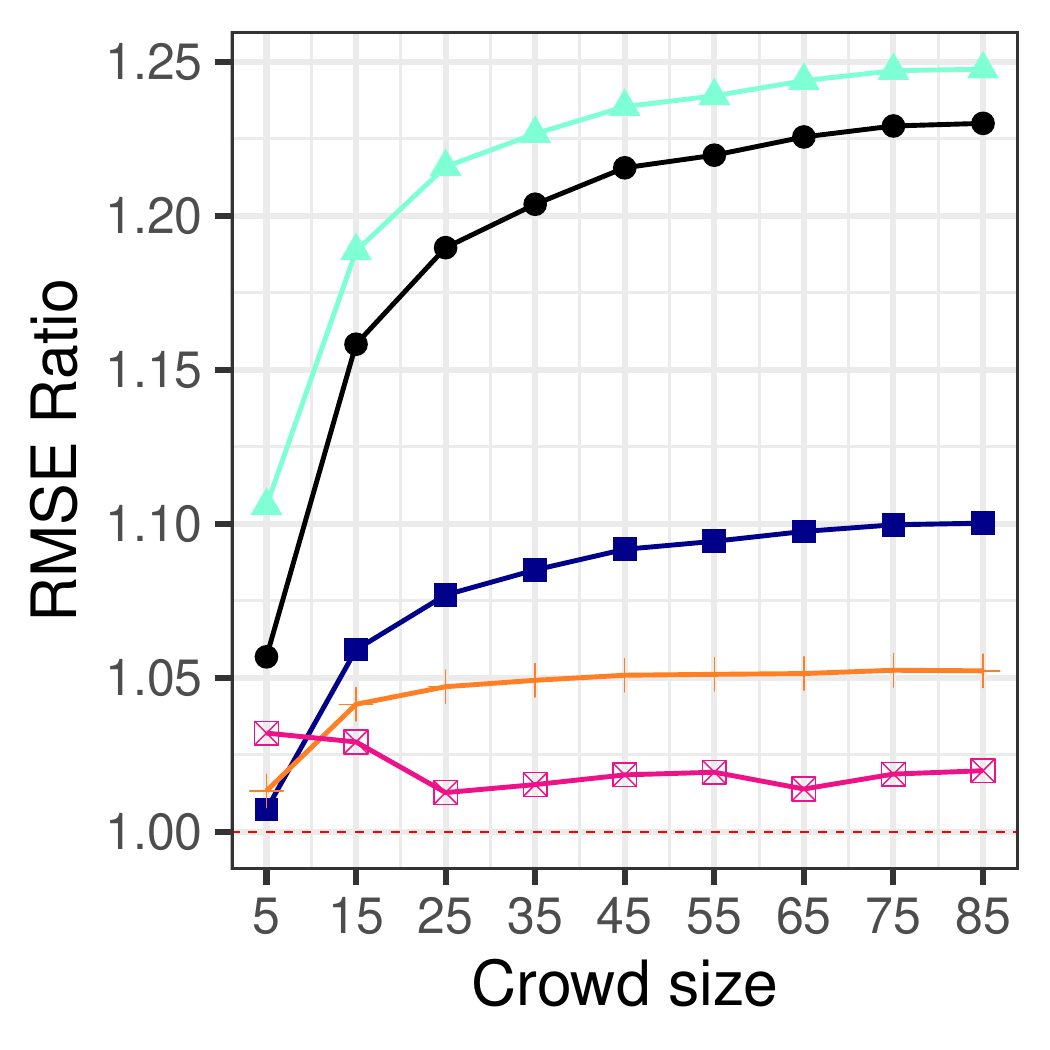} }}%
    \\
    \subfloat[\centering GK Diff. 3\label{fig:gk3}]{{\includegraphics[width=0.20\textwidth]{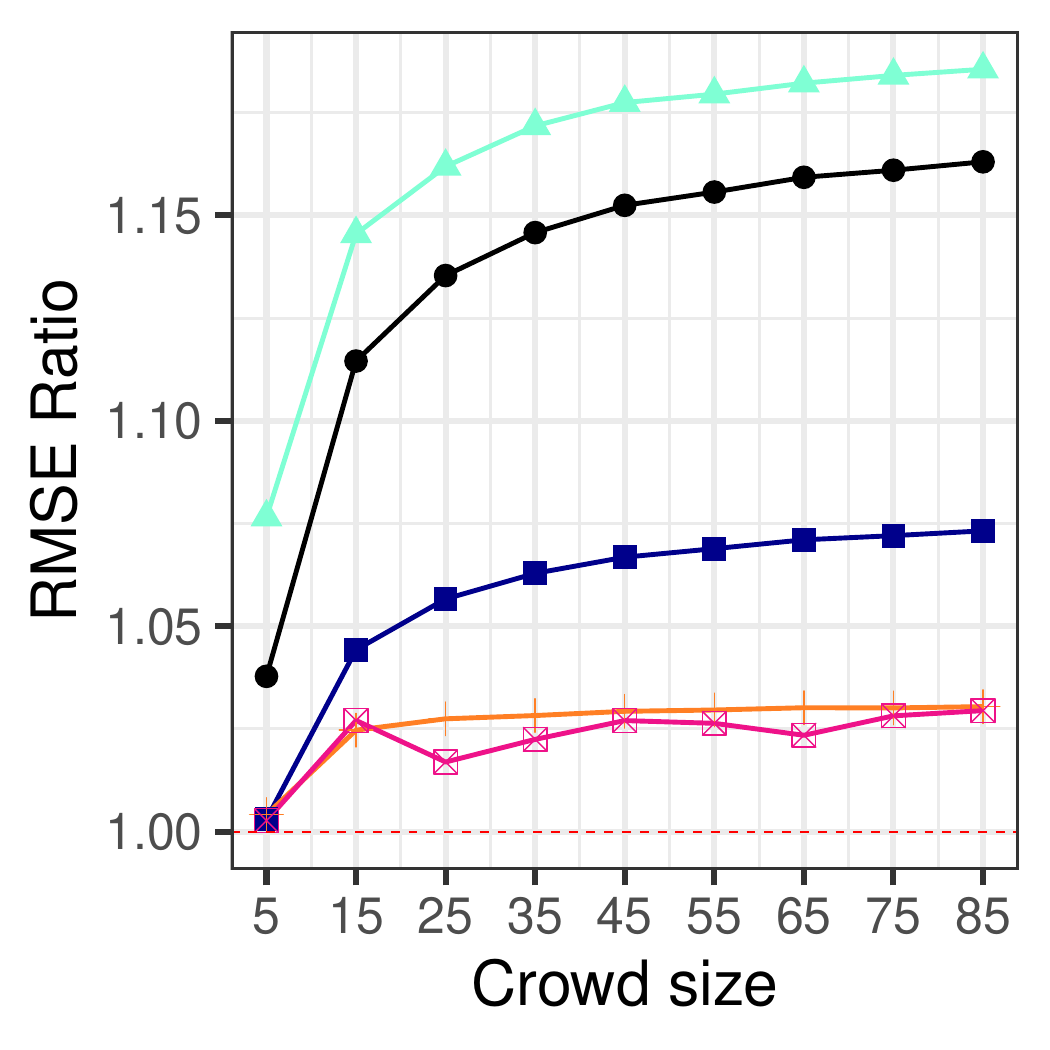} }}
    \quad
    \subfloat[\centering GK Diff. 4\label{fig:gk4}]{{\includegraphics[width=0.20\textwidth]{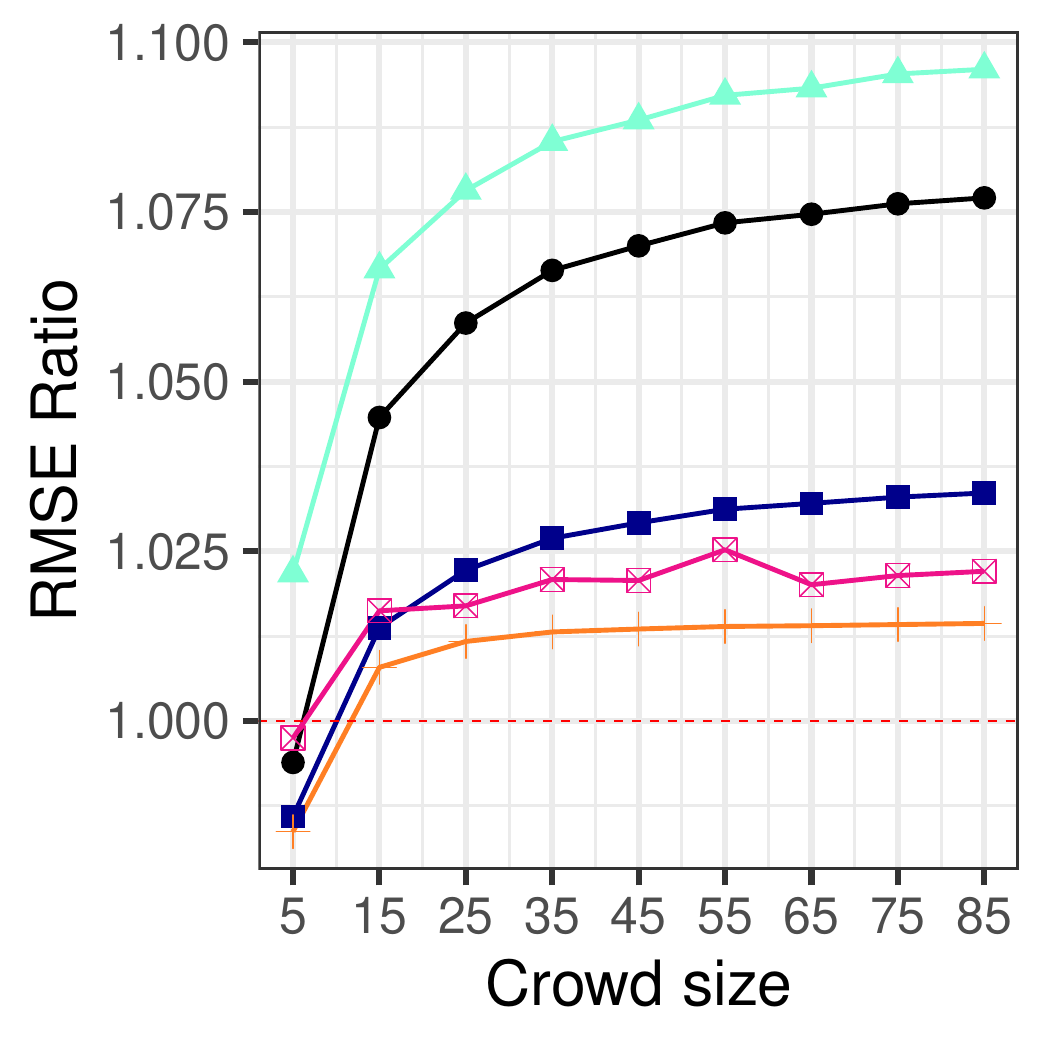} }}%
    \quad
    \subfloat[\centering GK Diff. 5\label{fig:gk5}]{{\includegraphics[width=0.20\textwidth]{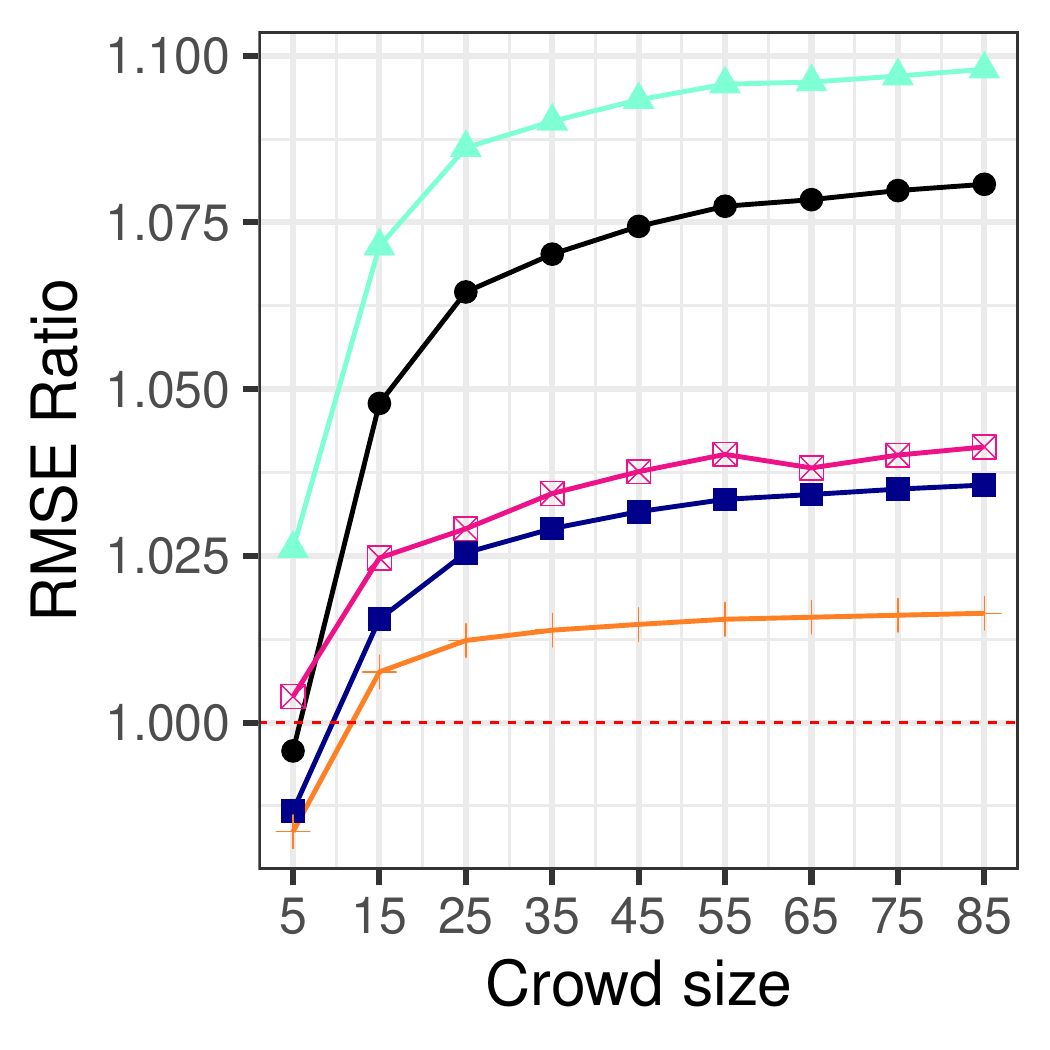} }}%
    \quad
    \subfloat[\centering NCAA Round of 16 \label{fig:ncaa16}]
    {{\includegraphics[width=0.20\textwidth]{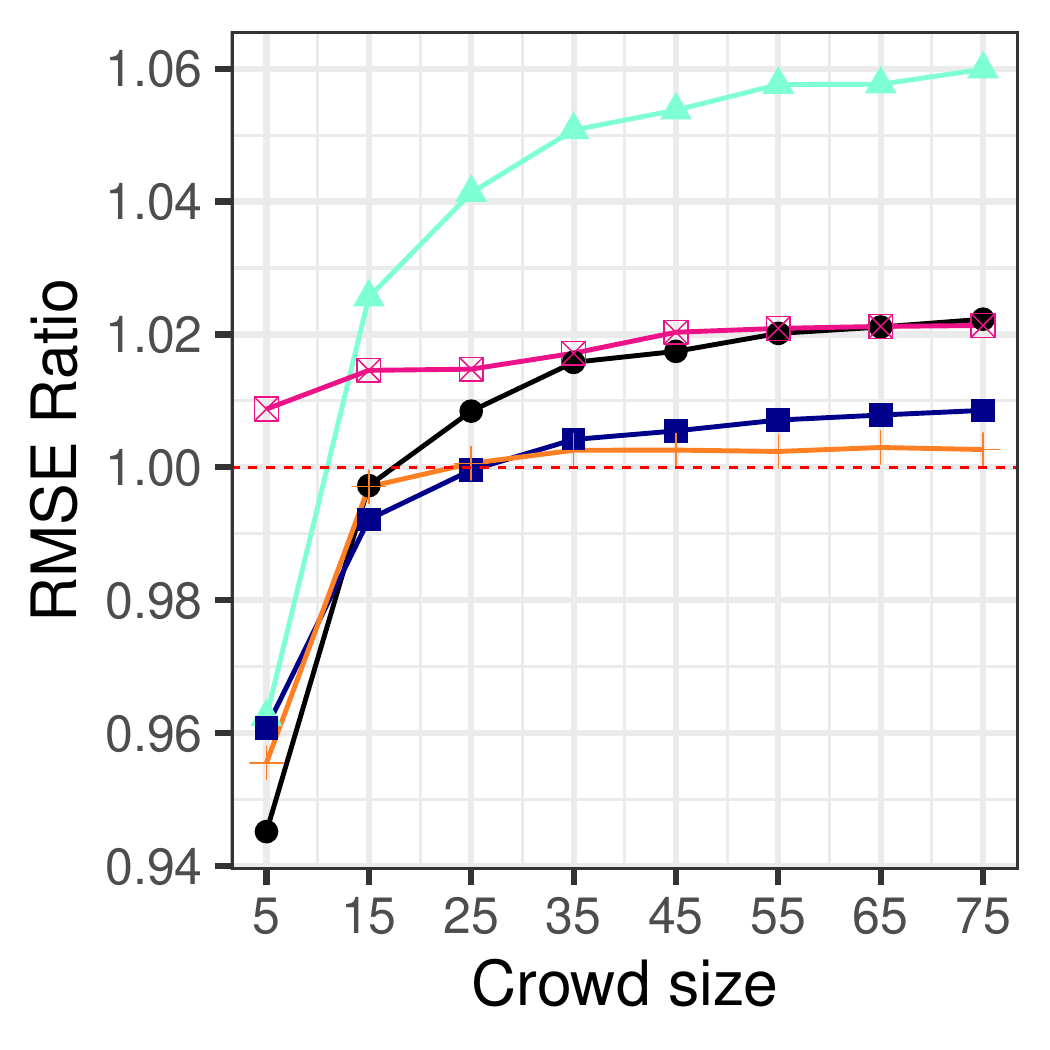} }}%
    \\
    \subfloat[\centering NCAA Round of 64 \label{fig:ncaa64}]
    {{\includegraphics[width=0.20\textwidth]{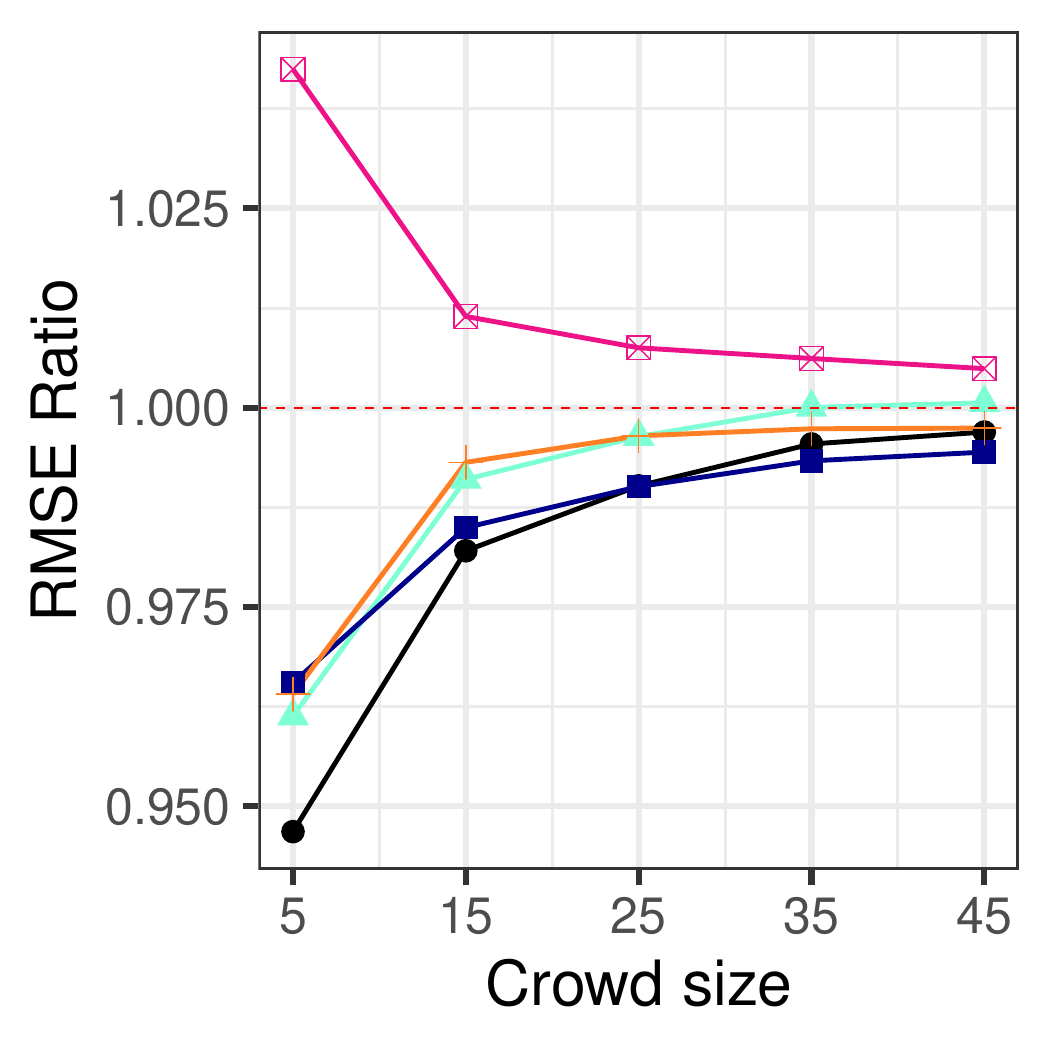} }}%
    \quad
    \subfloat[\centering Coins Sym. \label{fig:s}]{{\includegraphics[width=0.20\textwidth]{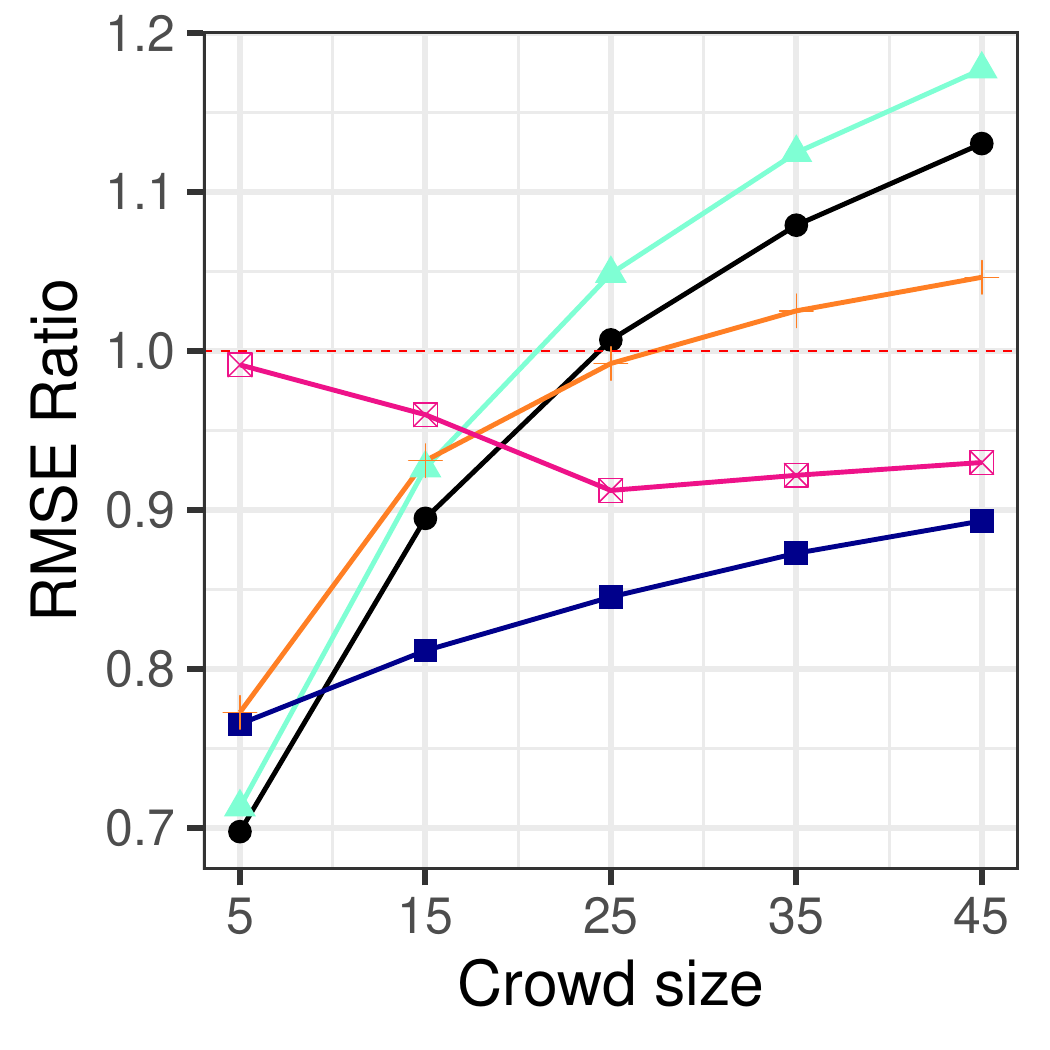} }}%
    \quad
    \subfloat[\centering Coins Nested Sym. \label{fig:ns}]{{\includegraphics[width=0.20\textwidth]{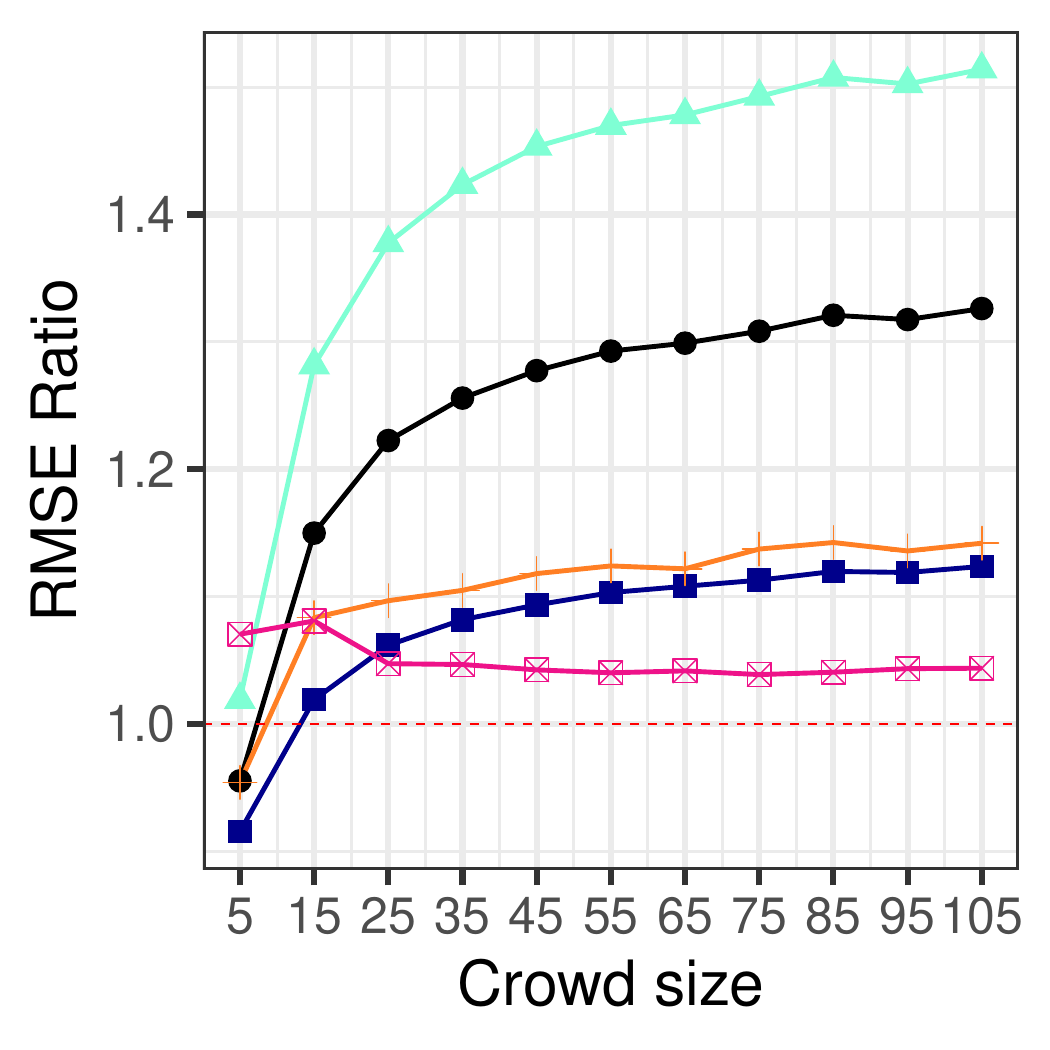} }}%
    \quad
    \subfloat[\centering Coins Nested \label{fig:n}]{{\includegraphics[width=0.20\textwidth]{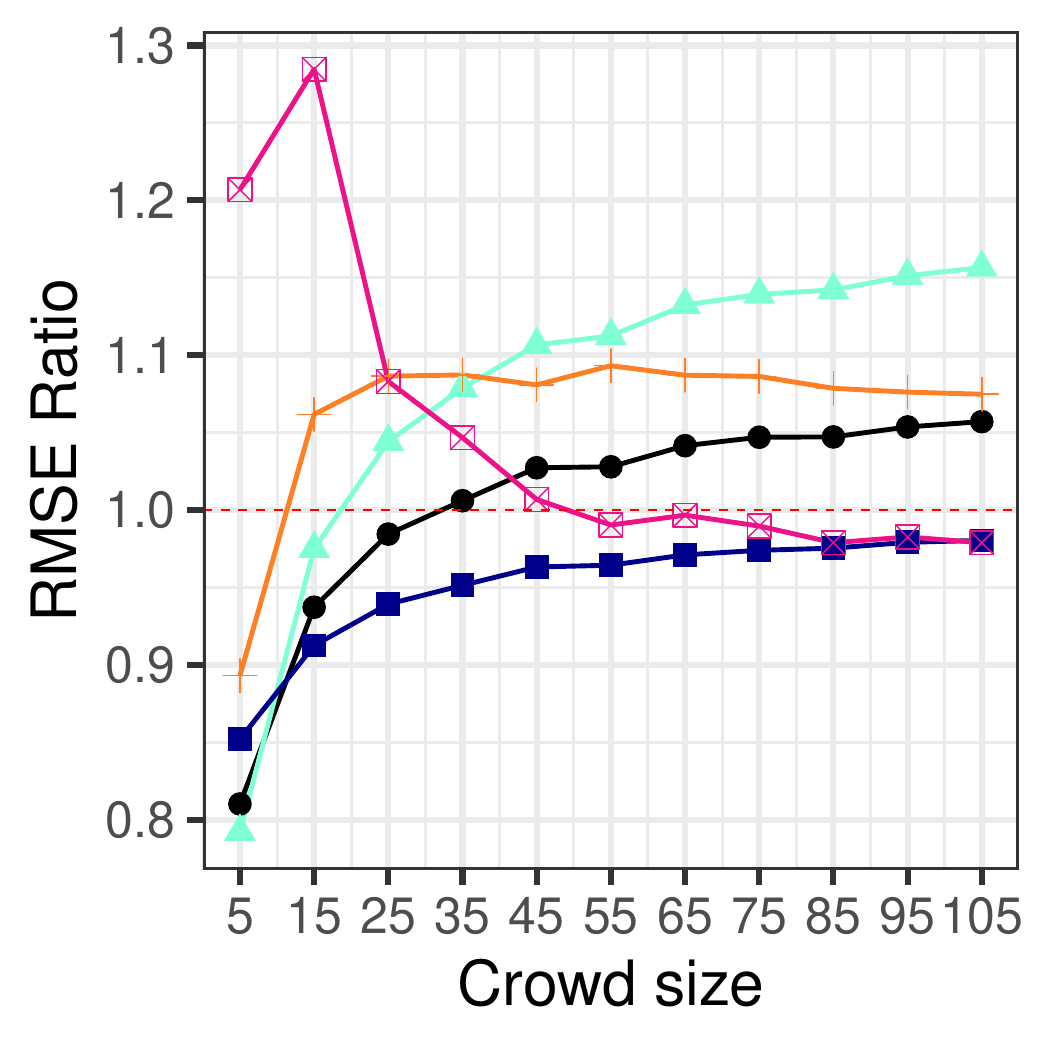} }}%

    \caption{Bootstrap root mean squared error divided by root mean squared error of Neutral Pivoting. The \textcolor{red}{red} dashed line is set at one. Legend: \textcolor{blue}{MP} ($\blacksquare$), \textcolor{YellowOrange}{KW}($+$), \textcolor{magenta}{SO} (\usebox{\mytikzbox}), \textcolor{black}{Simple Average} ($\CIRCLE$), \textcolor{SkyBlue}{Median} ($\blacktriangle$)}%
    \label{fig:bootstrap}%
\end{figure}

Figure \ref{fig:bootstrap} plots the average ratio of competing methods' RMSE divided by the neutral pivot's RMSE for each method at each crowd size.

\end{appendix}

% References here (outcomment the appropriate case)

% CASE 1: BiBTeX used to constantly update the references
%   (while the paper is being written).
%\bibliographystyle{informs2014} % outcomment this and next line in Case 1
%\bibliography{<your bib file(s)>} % if more than one, comma separated

% CASE 2: BiBTeX used to generate mypaper.bbl (to be further fine tuned)
%\input{mypaper.bbl} % outcomment this line in Case 2

%If you don't use BiBTex, you can manually itemize references as shown below.
\bibliographystyle{informs2014} % Style BST file (imsart-number.bst or imsart-nameyear.bst)
\bibliography{reference}  

\begin{thebibliography}{8}
\providecommand{\natexlab}[1]{#1}
\providecommand{\url}[1]{\texttt{#1}}
\providecommand{\urlprefix}{URL }

\bibitem[{Armstrong(2001)}]{hillier_combining_2001}
Armstrong JS (2001) Combining {Forecasts}. Hillier FS, Armstrong JS, eds., \emph{Principles of {Forecasting}}, volume~30, 417--439 (Boston, MA: Springer US), ISBN 9780792374015 9780306476303, \urlprefix\url{http://dx.doi.org/10.1007/978-0-306-47630-3_19}.

\bibitem[{Jose \protect\BIBand{} Winkler(2008)}]{jose_simple_2008}
Jose VRR, Winkler RL (2008) Simple robust averages of forecasts: {Some} empirical results. \emph{International Journal of Forecasting} 24(1):163--169, ISSN 01692070, \urlprefix\url{http://dx.doi.org/10.1016/j.ijforecast.2007.06.001}.

\bibitem[{Martinie et~al.(2020)Martinie, Wilkening, \protect\BIBand{} Howe}]{mart}
Martinie M, Wilkening T, Howe PDL (2020) Using meta-predictions to identify experts in the crowd when past performance is unknown. \emph{PLOS ONE} 15(4):e0232058, ISSN 1932-6203, \urlprefix\url{http://dx.doi.org/10.1371/journal.pone.0232058}.

\bibitem[{Palley \protect\BIBand{} Satopää(2023)}]{kw}
Palley AB, Satopää VA (2023) Boosting the wisdom of crowds within a single judgment problem: Weighted averaging based on peer predictions. \emph{Management Science} 69(9):5128–5146, ISSN 0025-1909, 1526-5501, \urlprefix\url{http://dx.doi.org/10.1287/mnsc.2022.4648}.

\bibitem[{Palley \protect\BIBand{} Soll(2019)}]{mp}
Palley AB, Soll JB (2019) Extracting the wisdom of crowds when information is shared. \emph{Management Science} 65(5):2291--2309, \urlprefix\url{http://dx.doi.org/10.1287/mnsc.2018.3047}.

\bibitem[{Peker(2023)}]{so}
Peker C (2023) Extracting the collective wisdom in probabilistic judgments. \emph{Theory and Decision} 94(3):467–501, ISSN 1573-7187, \urlprefix\url{http://dx.doi.org/10.1007/s11238-022-09899-4}.

\bibitem[{Stock \protect\BIBand{} Watson(2004)}]{stock_combination_2004}
Stock JH, Watson MW (2004) Combination forecasts of output growth in a seven‐country data set. \emph{Journal of Forecasting} 23(6):405--430, ISSN 0277-6693, 1099-131X, \urlprefix\url{http://dx.doi.org/10.1002/for.928}.

\bibitem[{Surowiecki(2005)}]{woc}
Surowiecki J (2005) \emph{The Wisdom of Crowds} (Anchor), ISBN 0385721706.

\end{thebibliography}

%%%%%%%%%%%%%%%%%
\end{document}